\documentclass[12pt]{article}
\usepackage{amsfonts,amsmath,amssymb,authblk,color,enumerate,float,graphicx,multicol,multirow,placeins,pifont,slashed,titlesec,xspace}
\usepackage[colorlinks=true,urlcolor=blue,anchorcolor=blue,citecolor=blue,filecolor=blue,linkcolor=blue,menucolor=blue,pagecolor=blue,linktocpage=true]{hyperref}
\usepackage[compress,numbers]{natbib}
\usepackage[labelfont=bf]{caption}
\usepackage[top=1in,bottom=1in,left=1in,right=1in,footskip=0.5in,headheight=0.5in,footnotesep=0.2in,marginparwidth=0in,marginparsep=0in]{geometry}
\usepackage{tikzfeynman}
\usepackage{aas_macros}
\usepackage{hyphenat}
\usepackage{enumitem}

\renewcommand{\arraystretch}{1.5}
\long\def\symbolfootnote[#1]#2{\begingroup%
\def\thefootnote{\fnsymbol{footnote}}\footnote[#1]{#2}\endgroup}

\newcommand{\abs}[1] {\big | #1 \big |}
\newcommand\notype[1]{\unskip}

\begin{document}
\begin{flushright}
\small YITP-SB-18-42
\end{flushright}

\vspace*{0.6cm}

\begin{center}
\pagenumbering{gobble}
{\LARGE \bf Structure Formation and Exotic Compact Objects in a Dissipative Dark Sector}
\bigskip\vspace{1cm}

{\large
Jae Hyeok Chang$^1$, Daniel Egana-Ugrinovic$^1$, \\
 Rouven Essig$^1$, and Chris Kouvaris$^{2}$ }
\\[7mm]
 {\small
 $^1$\,C. N. Yang Institute for Theoretical Physics, Stony Brook, NY 11794, USA \\
$^2$\,CP3-Origins, University of Southern Denmark, Campusvej 55, DK-5230 Odense M, Denmark
 }
 
\end{center}

%%%%%%%%%%%%%%%%%%%%%%%%%%%%%%%%%%%%
%%%%%%%%%%%%%%%%%%%%%%%%%%%%%%%%%%%% 
\vspace{4ex}
{\abstract
We present the complete history of structure formation in a simple dissipative dark-sector model.
The model has only two particles: a dark electron, which is a subdominant component of dark matter, and a dark photon.
Dark-electron perturbations grow from primordial overdensities, 
become non-linear, and form dense dark galaxies.
Brems\-strahlung cooling leads to fragmentation of the dark-electron halos into clumps that vary in size from a few to millions of solar masses, 
depending on the particle model parameters.
In particular, we show that asymmetric dark stars and black holes form within the Milky Way from the collapse of dark electrons.
These exotic compact objects may be detected and their properties measured at new high-precision astronomical observatories, 
giving insight into the particle nature of the dark sector without the requirement of non-gravitational interactions with the visible sector.
}

\newpage
\pagenumbering{arabic}
\setcounter{page}{2}
\tableofcontents
\newpage

%%%%%%%%%%%%%%%%%%%%%%%%%%%%%%%
%% Section break
%%%%%%%%%%%%%%%%%%%%%%%%%%%%%%%
\section{Introduction}
In spite of an extensive experimental program aiming to uncover the particle nature of dark matter, 
no dark sector particle has yet been discovered.
To this date, all the evidence for the existence of dark matter,  
remains purely gravitational in nature.
While many theories of dark matter predict non-gravitational interactions with the Standard Model and motivate a broad experimental program, 
it is possible that the dark sector is entirely secluded from the Standard Model.  

In the absence of a positive detection signal,
significant progress in the understanding of dark matter can still be made based uniquely on cosmological and astronomical observations,
which do not rely on non-gravitational interactions of the dark and visible sectors.
While on scales above $\sim 100 \, \textrm{kpc}$ cosmological observations are broadly consistent with dark matter being cold and collisionless (CDM),
astronomical observations 
of smaller scales may resolve interesting thermodynamic properties of the dark sector. 
Groundbreaking progress in the study of dark matter on small scales will be realized by a new generation of high-precision astronomical observatories such as LIGO~\cite{Abbott:2007kv,TheLIGOScientific:2014jea}, Gaia~\cite{Prusti:2016bjo} and LSST~\cite{Ivezic:2008fe,Abell:2009aa,lsstSRD}. 

In this scenario, the main challenge for particle physicists is to turn the 
experimental program of high-precision observatories into a theory program for particle dark matter.
One task is to start with particle physics models and provide calculable predictions for the dark-sector structure on small scales,
so that in the event of an observation departing from the CDM paradigm, 
one could pinpoint the underlying dark-sector model.
Work in this direction has been mostly concentrated in the study of  primordial black holes~\cite{Carr:2016drx} and axion stars~\cite{Tkachev:1991ka,Hogan:1988mp,Schive:2014dra,Eby:2016cnq}.
These are interesting objects on their own, 
but they have a formation history that shares no resemblance with the one of baryons,
which is the only example of structure formation that departs from CDM and we know for sure was realized.
However,
if dark matter is anything like the baryonic sector, the problem of understanding structure formation becomes formidable, since 
baryons form structure by the linear growth of perturbations that later undergo 
a rather complicated non-linear evolution that is accompanied by cooling. 
Numerical simulations provide insight into the non-linear regime, 
but they are time-consuming and computationally expensive, so they are not necessarily the best approach in an initial stage 
of theory exploration.

Efforts in understanding small-scale structure in dissipative dark-sector models are already underway,
and generically rely on models that mimic the Standard Model to benefit from the intuition of baryonic structure formation 
\cite{CyrRacine:2012fz,McCullough:2013jma,Fan:2013bea,Fischler:2014jda,Foot:2014uba,Foot:2016wvj,Agrawal:2017rvu,Cyr-Racine:2013fsa,Foot:2017dgx,Foot:2018qpw,Foot:2014mia,Buckley:2017ttd,Boddy:2016bbu}.
However,
the history of structure formation in the Standard Model is complicated, 
so in any model that mimics the baryonic sector it is challenging to get calculable predictions.

In this work,
we propose instead to start with the study of a simple dark-sector model that forms interesting small-scale structure via cooling, 
as baryons do,
but that is stripped off the details and complications of the baryonic sector,
most notably the existence of bound states.\footnote{Differently the models in~\cite{Wise:2014jva,Gresham:2017cvl,Gresham:2018anj}.}
The model contains only two particles: a dark electron,
which is responsible for forming astronomical objects and is a subdominant component of dark matter,
and a massive dark photon, 
which mediates dark-electron self-interactions and leads to the cooling of dark-electrons via bremsstrahlung.
Non-gravitational interactions of the dark sector with the Standard Model are \textit{not needed} in this model.
A complete dark-matter model would likely contain more fields and complicated interactions,
but we will show that this simple model already provides valuable insight for understanding small-scale structure formation in the dark sector.
Very much like the baryonic sector,
our dark-electron sector is a subdominant component of the matter content in the Universe and
is asymmetric in nature~\cite{Nussinov:1985xr,Barr:1990ca,Cosme:2005sb,Gudnason:2006yj,
Kaplan:2009ag,Zurek:2013wia,Petraki:2013wwa}  (so there is no significant abundance of dark-positrons).
It also forms cosmological and astronomical structure as baryons do, 
starting from primordial overdensities that eventually decouple from the Hubble flow
and become compact, gravitationally-bound objects.
However, in contrast to the case of baryons, 
the simplicity of our dark-electron sector
allows a straightforward description of its structure formation history
and to effectively calculate the typical mass and size of the final astronomical objects as a function of the particle model parameters.
In particular, the absence of bound states significantly simplifies the analysis of cooling with respect to the case of baryons.

Despite the simplicity of the model,
dark electrons form very interesting structure on sub-galactic scales. 
We find that due to cooling a galactic halo consisting of dark-electrons fragments, 
as the baryonic galactic halo does,
and forms a variety of exotic compact objects,
ranging from compact solar-mass-sized dark-electron ``asymmetric stars'' 
\footnote{Asymmetric dark stars as gravitationally stable compact objects were first proposed and their properties studied in~\cite{Kouvaris:2015rea} for fermionic dark matter and in~\cite{Eby:2015hsq} for bosonic dark matter.}
 to large supermassive black holes. 
Our final results are summarized in Fig.~\ref{fig:five},
where we present the mass and compactness of the exotic compact objects formed by halo fragmentation, 
as a function of the particle-model parameters. 
These exotic compact objects could be extremely abundant,
even if dark electrons are only a small subdominant component of dark-matter.
In particular,
assuming that the dark-electron sector corresponds to $1\%$ of dark-matter,
we find that there could be as much as one solar-mass-sized asymmetric star consisting of dark electrons 
for every ten baryonic stars within the Milky Way.

We organize this paper as follows. 
In Section~\ref{sec:two}, we present the simplified dark-sector-particle model, discuss cosmological abundances and generic bounds on its parameter space.
In Section~\ref{sec:three}, we study the growth of dark-electron overdensities in the linear regime,
starting from the primordial perturbations contained in a scale-invariant power spectrum,
and we identify the parameter space leading to the formation of interesting structure departing from the CDM paradigm.
In Section~\ref{sec:four}, we study the growth of perturbations in the non-linear regime on galactic scales,
study the fragmentation of the halo due to cooling using a Jeans analysis,
and calculate the mass of the smallest dark-electron fragments, which correspond to exotic compact objects. 
We study the stability of these objects and calculate their compactness. 
We also discuss the limitations of our analysis; in particular, we do not include angular momentum in the study of dark-electron halo collapse.
In Section~\ref{sec:five}, we briefly comment on the experimental signatures that could be pursued at high-precision astronomical observatories.
We leave technical details for appendices.

%%%%%%%%%%%%%%%%%%%%%%%%%%%%%%%
%%%%%%%%%%%%%%%%%%%%%%%%%%%%%%%

%%%%%%%%%%%%%%%%%%%%%%%%%%%%%%%
%% Section break
%%%%%%%%%%%%%%%%%%%%%%%%%%%%%%%

%%%%%%%%%%%%%%%%%%%%%%%%%%%%%%%
%%%%%%%%%%%%%%%%%%%%%%%%%%%%%%%
\section{A simple dissipative dark sector model}
\label{sec:two}
We consider a model that is an Abelian gauge theory containing a Dirac fermion field ${\Psi}_{e_D} $ and a gauge boson $A_{\mu}$,
which we refer to as the dark-electron and dark photon fields. 
Spontaneous breaking of the gauge symmetry gives a mass to the dark photon. 
The corresponding Lagrangian is
%%boe%%
\begin{eqnarray}
& & 
\mathcal{L} \supset
i \,\bar{\Psi}_{e_D}  {\gamma}^\mu D_\mu {\Psi}_{e_D} 
-
m_{e_D}
\bar{\Psi}_{e_D} {\Psi}_{e_D} 
-
 \frac{1}{4}
F_{\mu \nu}F^{\mu \nu}
+
m_{\gamma_D}^2 A_\mu A^\mu
\quad ,
\label{eq:lagrangian}
 \end{eqnarray}
where $m_{e_D}, m_{\gamma_D}$ are the dark-electron and dark photon-masses, respectively, 
and the gauge coupling is defined with the normalization
\begin{equation}
D_\mu=\partial_\mu-i q_e g_D A_\mu \quad .
\end{equation}
%%eoe%%
In what follows and without loss of generality we take the dark-electron charge to be $q_e=1$.
We also define a dark fine-structure constant as $\alpha_D\equiv g_D^2/(4\pi)$.
Kinetic mixing of the dark photon with the Standard Model photon \textit{is not necessary} in what follows.

The objective in this paper is to determine if this model forms astronomical structure in analogy to %as 
the Standard Model, but consisting of gravitationally bound objects composed of dark electrons and dark photons instead of baryons.
Since dark electrons may cool via Brems\-strahlung, which favors the formation of compact astronomical objects,
we will be mostly interested in structure consisting of dark electrons rather than dark photons. 
In addition, we limit ourselves to the case $m_{e_D} \gg m_{\gamma_D}$ to allow for Brems\-strahlung even when dark electrons 
are non-relativistic, 
and we take the fine structure constant to be large to allow for efficient cooling.
In particular, we will see that the range $10^{-3} \lesssim \alpha_D \lesssim 10^{-1}$ will efficiently produce exotic compact objects.

\subsection{Cosmological abundances}

To avoid complications arising from bound states and dark-electron-dark-positron annihilations during structure formation, 
we impose that the dark sector is asymmetric and require the cosmological abundance of dark positrons to be negligible.
Since dark-electron number is conserved by the interactions in Eq.~\eqref{eq:lagrangian},
additional interactions breaking dark-electron number and CP are needed in order for the asymmetry to be created.
In addition, since at high temperatures the gauge symmetry is restored,
any charge asymmetry must be generated at temperatures $v_H \sim (g_D/q_H) \, m_{\gamma_D}$, 
where $v_H$ is the Higgs condensate breaking the gauge symmetry and $q_H$ the Higgs gauge charge. 
For brevity, in this work we do not specify the full mechanism leading to our effective theory nor the Higgs sector breaking the dark U(1) gauge symmetry,
but we refer the reader to~\cite{Petraki:2014uza} for an example on how to generate the electric charge asymmetry and on the details of the Higgs sector.
Here we simply parametrize the asymmetry by the ratio $f$ of the background dark-electron energy density $\rho^{e_D}_0$ to the background dark-matter energy density ${\rho^{\textrm{DM}}_0}$,
\begin{equation}
f 
\equiv
\frac{\rho^{e_D}_0}
{\rho^{\textrm{DM}}_0} 
\quad .
\label{eq:DMfraction}
\end{equation}
For the dark-positron abundance to be negligible at present times, 
dark positrons must efficiently annihilate with dark electrons in the early Universe. 
For concreteness, we require their relic abundance to be less than a percent of the asymmetric dark-electron abundance. 
This condition sets a minimal value for the fine structure constant to ensure efficient annihilations, 
given by~\cite{Petraki:2014uza,Graesser:2011wi} 
\begin{equation}
\alpha_D
\geq
4.6\,
\times
10^{-7}
\,
\bigg[
\frac{m_{e_D}}
{1\, \textrm{MeV}}
\bigg]
\,
\bigg[
\frac{10^{-2}}
{f}
\bigg]^{1/2}
\bigg[
\frac{T_{e_D}|_{e_D \, {\textrm{dec}}}}
{T_{\textrm{SM}}|_{e_D \, {\textrm{dec}}}}
\bigg]
^{1/2}
\quad ,
\label{eq:symmetricabundance0}
\end{equation}
where $T_{e_D}|_{e_D \, {\textrm{dec}}}$ and $T_{\textrm{SM}}|_{e_D \, {\textrm{dec}}}$ are the dark-electron and Standard-Model temperatures at dark electron-dark positron chemical decoupling. 
The condition \eqref{eq:symmetricabundance0} may be equivalently rewritten as a condition on the dark electron temperature,
\begin{equation}
T_{e_D}|_{e_D \, {\textrm{dec}}}
\leq
4.6
\times
10^8
\,
\bigg[
\frac{\alpha_D }
{10^{-2}}
\bigg]^2
\,
\bigg[
\frac{1\, \textrm{MeV}}
{m_{e_D}}
\bigg]^2
\bigg[
\frac{f}
{10^{-2}}
\bigg]
\,
T_{\textrm{SM}}|_{e_D \, {\textrm{dec}}}
\quad .
\label{eq:symmetricabundance}
\end{equation}

The abundance of dark photons is also determined by the annihilation between dark-electrons and dark-positrons, 
which sets the temperature of dark-photon chemical decoupling.
Since $m_{e_D} \gg m_{\gamma_D}$ and since our fine structure constant is large, dark photons come out of chemical equilibrium while relativistic in our minimal model.
Dark photons are then hot  relics and their abundance after they become non-relativistic is given by  
\begin{equation}
\rho^{\gamma_D}_0
=
s Y_{\gamma_D} m_{\gamma_D}
\label{eq:mgammadensity}
\quad ,
\end{equation}
where $s=(2\pi^2/45) g_{*S} T_{\textrm{SM}}^3$ is the entropy density  and 
\begin{equation}
Y_{\gamma_D}
=
s^{-1}|_{\gamma_D \, {\textrm{dec}}} \,
\frac{2 \zeta(3)}{\pi^2} T_{\gamma_D}^3 |_{\gamma_D \, {\textrm{dec}}}  \quad ,
\end{equation}
where $\zeta(3)\simeq 1.2$ and $s|_{\gamma_D \, {\textrm{dec}}} $ and $T_{\gamma_D} |_{\gamma_D \, {\textrm{dec}}}$ are the entropy density and dark-photon temperature at chemical decoupling.
Eq.~\eqref{eq:mgammadensity} sets a maximal dark-photon temperature at chemical decoupling to avoid overclosure, 
given by
\begin{equation}
T_{\gamma_D}|_{\gamma_D \, {\textrm{dec}}}
\leq
0.2 
\bigg[
\frac{g_{*S}|_{\gamma_D \, {\textrm{dec}}}}
{10}
\bigg]^{1/3}
\bigg[
\frac{1\, \textrm{keV}}
{m_{\gamma_D}}
\bigg]^{1/3}
\,\,
T_{\textrm{SM}}|_{\gamma_D \, {\textrm{dec}}} 
\quad ,
\label{eq:overclosure}
\end{equation}
where $T_{\textrm{SM}}|_{\gamma_D \, {\textrm{dec}}}$ is the Standard-Model temperature and $g_{*S}|_{\gamma_D \, {\textrm{dec}}}$ the number of effective degrees of freedom, at dark-photon chemical decoupling.

In addition, 
since the formation of exotic compact objects 
requires a light dark photon for efficient cooling through Bremsstrahlung,
dark photons may be relativistic at nucleosynthesis, in which case their temperature is strongly constrained  from the effective number of relativistic species, $N_{\textrm{eff}}$~\cite{Cyburt:2015mya}.
To allow for light dark photons we require their temperature at nucleosynthesis to be 
\begin{equation}
T_{\gamma_D}
|_{{\textrm{BBN}}} 
\leq 0.5 \, T_\mathrm{{\textrm{SM}}}
|_{{\textrm{BBN}}} \quad. 
\label{eq:neff}
\end{equation}
Note that for a dark-photon mass $m_{\gamma_D} \gtrsim 8 \, \textrm{eV}$ and neglecting $\mathcal{O}(1)$ numbers coming from factors of $g_{*S}^{1/3}$,
the overclosure limit, Eq.~\eqref{eq:overclosure}, is stronger than the limit from  the effective number of relativistic species, 
Eq.~\eqref{eq:neff}.
More generally, we impose that the dark-photon temperature satisfies the strongest of the limits Eqns.~\eqref{eq:overclosure} or \eqref{eq:neff}. 
The difference in temperature between our dark sector and the Standard Model may for instance come from an asymmetric reheating scenario~\cite{Adshead:2016xxj}. 

In this work, we assume that the total dark-matter abundance consists of dark electrons, dark photons, 
and other cold or warm dark-sector particles that we do not specify. 
Since dark electrons self-interact via the dark-photon force and we allow for a large fine structure constant,
the dark-electron sector is generically strongly self-interacting. % and may not be the dominant dark matter component.
To avoid bounds from halo shapes and the bullet cluster~\cite{Tulin:2017ara,Rocha:2012jg}, 
we impose that dark electrons are no more than ten percent of the total dark matter, $f  < 10 \%$~\cite{Fan:2013yva}. 

In a minimal setup, dark photons could in principle be the dominant component of dark matter,
as long as their mass satisfies $m_{\gamma_D} \geq 0.3\, \textrm{keV}$ to avoid washing out small-scale structure via free-streaming~\cite{Viel:2005qj}
(we discuss this in more detail in Appendix~\ref{app:one}).
However, assessing the viability of this possibility would require a more careful analysis of structure formation in the dark-photon sector, 
in addition to the study of structure formation in the dark-electron sector. 
This is beyond the scope of this work,
so in what follows we assume that dark photons are also a subdominant component of dark matter, 
and that the dominant dark matter species is cold and non interacting. 
Dark photons are clearly subdominant if the dark-photon temperature is a factor of a few below the overclosure bound in Eq.~\eqref{eq:overclosure}. 

We now move on to study structure formation in the dark-electron sector. 
The growth of dark-electron and dark-photon perturbations starts from the small primordial overdensities from inflation.
In the next section, we discuss the evolution of these perturbations in the initial stage of gravitational collapse using linear perturbation theory.

\section{Growth of perturbations in the linear regime}
\label{sec:three}
In the first stage of gravitational collapse, 
the matter perturbations are smaller than the background density, $\delta \equiv \delta\rho/\rho \leq 1$. 
In this stage,
a linear analysis may be performed to study the evolution of dark electron overdensities and to determine the conditions for these perturbations to grow. 
If the conditions for linear growth are fulfilled, 
dark-electron overdensities may eventually become non-linear, 
decouple from the Hubble flow and form self-gravitating objects. 

The overall size of a perturbation in the linear regime at some redshift is determined entirely by the gas thermodynamics, gravity, and by the initial conditions for the perturbations.
To make immediate contact with our simplified model, 
we start by summarizing the basic thermodynamic properties of the dark-electron gas in Section~\ref{sec:therodynamics}.
In Section~\ref{sec:lineardynamics}, we discuss the conditions for linear growth of the dark-electron perturbations.
Finally, in Section~\ref{sec:transition}, we discuss the initial conditions of the perturbations and the transition into the non-linear regime.

\subsection{The three thermodynamic regimes of the dark-electron gas}
\label{sec:therodynamics}
The growth of matter perturbations depends on the interplay between gravity and the restoring pressure support provided by sound waves.
The dark-photon and dark-electron gas may be in different thermodynamic regimes,
each one with a characteristic pressure and speed of sound. 
The transition between the different regimes depends on the dark-electron and dark-photon mean free paths,
which are given by
%%%
\begin{equation}
\ell_{e_D}
=
1/(\sigma_M n_{e_D})
\quad ,
\quad 
\ell_{\gamma_D}^{C}
=
1/(\sigma_C n_{e_D})
\quad ,
\label{eq:meanfreepaths}
\end{equation}
%%%
where $\sigma_M$ and $\sigma_C$ are the M\"oller and Compton scattering cross sections,
and $n_{e_D}\equiv \rho_{e_D}/m_{e_D}$ is the dark-electron number density.
The Compton and M\"oller scattering cross sections in the non-relativistic limit are given by
%%%
\begin{equation}
\sigma_C
=
\zeta
\frac{8\pi}{3} 
\frac{\alpha_D^2}{m_{e_D}^2}
\bigg[
1
+
\mathcal{O}
\bigg(
\frac{m_{\gamma_D}^2 }
{m_{e_D}^2 }
,
{v^2_{e_D}}
\bigg)
\bigg]
\label{eq:compton}
\quad ,
\end{equation}
\begin{equation}
 \sigma_M
 =
4\pi
\frac{\alpha_D^2m_{e_D}^2}
{m_{\gamma_D}^4}
\bigg[
1
+
\mathcal{O}
\bigg(
\frac{m_{e_D}^2 {v^2_{e_D}}}
{m_{\gamma_D}^2 }
,
{v^2_{e_D},
\frac{\alpha_D m_{\gamma_D}}
{m_{e_D}v_{e_D}^2}
}
\bigg)
\bigg]
\quad ,
\label{eq:sigmaR}
\end{equation}
%%%
where $v_{e_D} \simeq \sqrt{3T_{e_D}/m_{e_D}}$ is the velocity of the non-relativistic dark electrons,
and $\zeta=1$ or $\zeta=2/3$ for the polarization-averaged cross section of  relativistic or non-relativistic dark photons, respectively. 
The Compton scattering cross section has corrections of order $m_{\gamma_D}^2/m_{e_D}^2$ that we neglect. 
For the M\"oller cross section,
we work in the ``contact interaction'' limit where the cross section is velocity independent,  
and we neglect non-perturbative corrections that arise for  $\alpha_D m_{\gamma_D} \geq  m_{e_D}v_{e_D}^2 \simeq T_{e_D}$,
which are only logarithmic in the expansion factor $\sim \log\big(
\alpha_D m_{\gamma_D}/T_{e_D}\big)$~\cite{PhysRevE.70.056405,Feng:2009hw} and do not significantly affect the conclusions of this section.
Comparing Eqns.~\eqref{eq:compton} and~\eqref{eq:sigmaR},
we see that the M\"oller scattering cross section is enhanced with respect to Compton by a factor of $m_{e_D}^4/m_{\gamma_D}^4$.  
As a consequence, 
the dark-electron mean free path is much smaller than the dark-photon mean free path,  
$\ell_{e_D}\ll \ell_{\gamma_D}$.
Subject to this hierarchy between the mean free paths, 
there are three possible thermodynamic regimes for the dark electron and dark-photon gas, 
which are set by comparing the proper or physical length scale $\lambda_P$ of any particular perturbation we wish to study with $\ell_{e_D}$ and $\ell_{\gamma_D}$.
The three thermodynamic regimes are:
\begin{eqnarray*}
& \ell_{\gamma_D} \gg \ell_{e_D} \gg \lambda_P  &  \textrm{collisionless dark electron regime.} 
\\ 
& \ell_{\gamma_D} \gg \lambda_P \gg \ell_{e_D} &    \textrm{self-interacting $e_D$ gas. $e_D$ and $\gamma_D$ decoupled.}   
\\
&\lambda_P  \gg \ell_{\gamma_D} \gg \ell_{e_D} &   \textrm{tightly coupled $e_D-\gamma_D$ gas.}
\end{eqnarray*}
The collisionless dark-electron regime $\ell_{e_D} \gg \lambda_P$ is rather trivial. 
In this regime, the dark electrons \textit{and} dark photons are kinetically decoupled on the scale of the perturbation, they have a vanishing speed of sound,
and behave as CDM. 

In the self-interacting dark-electron gas regime\footnote{
To ensure that dark electrons efficiently self-interact one must also check that the typical collision time between two dark electrons $\tau_c= \ell_{e_D}/v_{e_D}$ is shorter
than a Hubble time, $\tau_c \simeq \ell_{e_D}/\sqrt{3T_{e_D}/m_{e_D}} < H^{-1}$. 
Since we are interested in a strongly interacting dark-electron sector with $\ell_{e_D}$ well below $H^{-1}$, 
the collision-time condition is easily satisfied, unless the dark-electron sector is extremely cold.
}, $\ell_{\gamma_D} \gg \lambda_P \gg \ell_{e_D}$, 
dark-electron self-interactions are common on the scale of the perturbation, 
but dark photons remain collisionless.
In this case, dark electrons behave as a collisional gas due to efficient 
M\"oller self-interactions,
but are decoupled from the background dark-photon gas.
The total dark-electron pressure has two contributions,
one from the kinetic-equilibrium pressure,
and the second from the dark-photon repulsive force, 
and it is given by~\cite{Kouvaris:2015rea}
%%%
\begin{equation}
P_{e_D}
=
n_{e_D} T_{e_D}
+
\frac{2 \pi \alpha_D n_{e_D}^2}{m_{\gamma_D}^2} \quad ,
\label{eq:pressure}
\end{equation}
%%%
where $T_{e_D}$ is the dark-electron gas temperature. 
The corresponding speed of sound is 
%%%
\begin{equation}
c_{s}^{e_D} 
=
\sqrt{
\frac{
T_{e_D}}
{m_{e_D}}
+
\frac{4\pi \alpha_D n_{e_D}}
{m_{e_D} m_{\gamma_D}^2}
}
\quad .
\label{eq:idealgascs}
\end{equation}
%%%

Finally, 
in the tightly coupled regime, 
both dark electrons and dark photons scatter efficiently within the scale of the perturbation, 
and must be treated as a unique comoving electron-photon gas. 
For adiabatic modes, they share a common speed of sound given by~\cite{Gorbunov:2011zzc}
%%%
\begin{equation}
c_{s}^{e_D\gamma_D} 
=
\bigg[
\frac{
(c_s^{\gamma_D})^2
 + 
R_{e\gamma} 
\,
(c_s^{e_D})^2
}
{1+R_{e\gamma}}
\bigg]^{1/2}
\quad ,
\label{eq:tightcs}
\end{equation}
%%%
where $c_s^{e_D}$ is given in Eq.~\eqref{eq:idealgascs}, 
while the dark-photon speed of sound, $c_s^{\gamma_D}$, is $\sqrt{1/3}$ or  $\sqrt{T_{\gamma_D}/m_{\gamma_D}}$ for relativistic and non-relativistic dark photons, respectively,
and $T_{\gamma_D}$ is the dark-photon temperature. 
$R_{e\gamma}$ is proportional to the ratio of the dark-electron to dark-photon energy densities 
%%%
\begin{equation}
R_{e\gamma}
=
\xi \frac{\rho_0^{e_D}}{\rho_0^{\gamma_D}} \quad ,
\label{eq:R}
\end{equation}
%%%
where $\xi=3/4~(1)$ for relativistic (non-relativistic) dark photons,
$\rho_0^{\gamma_D}$ is the energy density of the dark photons given in Eq.~\eqref{eq:mgammadensity}, 
and $\rho_0^{e_D}$ is the asymmetric dark-electron energy density. 
Note that for vanishing dark-photon abundances, 
the electron-photon speed of sound Eq.~\eqref{eq:tightcs} reduces to Eq.~\eqref{eq:idealgascs} as expected.

\begin{figure}[t!]
\begin{center}
\includegraphics[width=9cm]{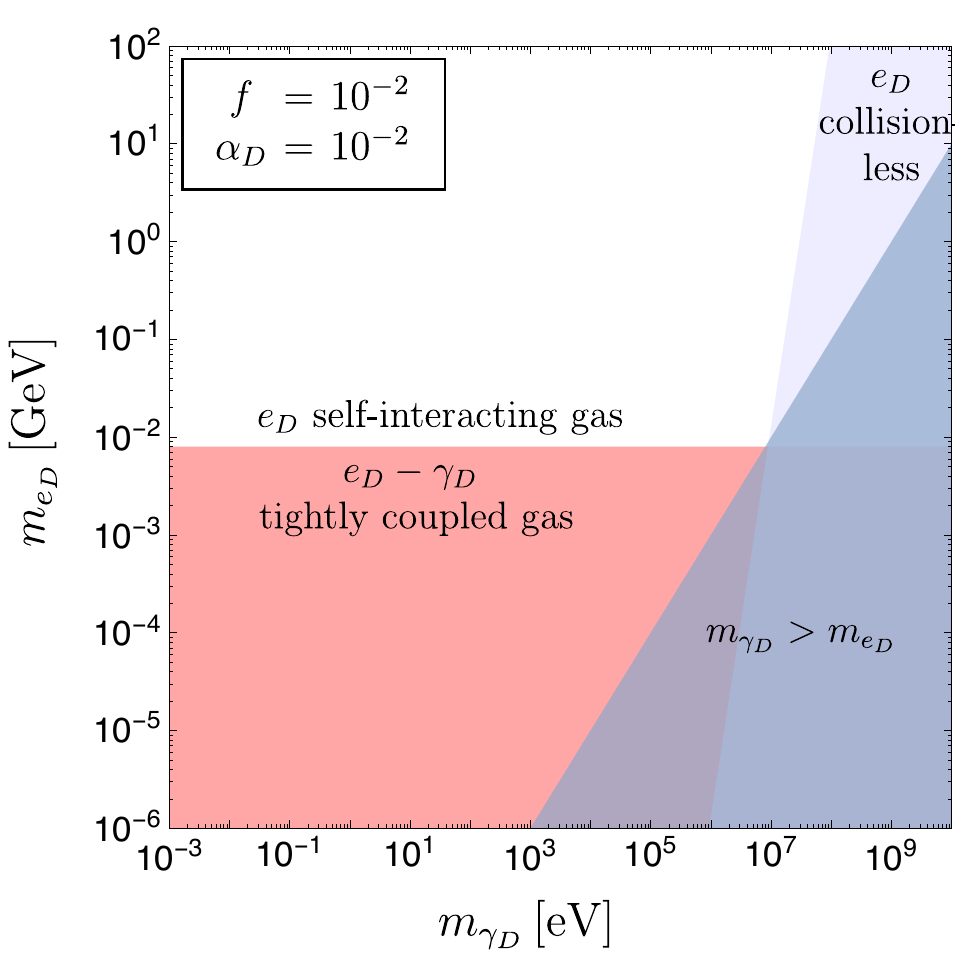}
\caption{
The three thermodynamic regimes of a dark-electron gas perturbation on a comoving scale that contains the Milky-Way mass, 
$\lambda_{\textrm{MW}}=4 \, \textrm{Mpc}$ (so that $\frac{\pi}{6}  \rho_0^{\textrm{DM}}|_{z=0} \, \lambda_{\textrm{MW}}^3 \simeq 10^{12} \, M_{\odot}$),
as a function of the dark-electron and dark-photon masses.
The boundary between the collisionless and the self-interacting regimes is given by 
$\ell_{e_D}=\lambda_{\textrm{MW}}/(1+z_{\textrm{eq}})$, 
while the boundary between the tightly-coupled and self-interacting regimes is given by $\ell_{\gamma_D}=\lambda_{\textrm{MW}}/(1+z_{\textrm{eq}})$,
where the dark-photon and dark-electron mean free paths $\ell_{\gamma_D,e_D}$ are given in Eqns.~\eqref{eq:meanfreepaths}, \eqref{eq:compton} and \eqref{eq:sigmaR}.
The redshift has been set to matter-radiation equality, $z_{\textrm{eq}}=3400$.
The dark fine-structure constant has been set to $\alpha_D=10^{-2}$,
while the dark electron background density has been set to be $f\equiv {\rho^{e_D}_0}/
{\rho^{\textrm{DM}}_0} = 1\%$ of the total dark-matter abundance.}
\label{fig:one}
\end{center}
\end{figure}

In Fig.~\ref{fig:one} we present the three different dark-electron thermodynamic regimes for a perturbation on the scale of the Milky Way, with current Lagrangian radius $\lambda_{\textrm{MW}}/2=2 \, \textrm{Mpc}$
(so $\frac{\pi}{6} \rho_0^{\textrm{DM}}|_{z=0} \, \lambda_{\textrm{MW}}^3 \simeq 10^{12} \, M_{\odot}$),
as a function of the dark-electron and dark-photon masses and at the onset of the linear growth of matter perturbations,
\textit{i.e.},
at matter-radiation equality $z_{\textrm{eq}}=3400$.
From Fig.~\ref{fig:one}, we see that for large dark-photon masses, 
which suppress M\"oller scattering,
dark electrons behave as a collisionless gas. 
%%%%%%
% FIGURE
%%%%%%
%%%%%%
% END OF FIGURE
%%%%%%
More generally,
at some redshift $z$, for any perturbation of a given current (or comoving) size $\lambda_C=(1+z)\lambda_P$,
there is a threshold value for the dark-photon mass above which dark electrons are collisionless. 
This threshold mass is obtained by setting $\ell_{e_D}=\lambda_P$ and using Eqns.~\eqref{eq:meanfreepaths}, \eqref{eq:sigmaR}, 
and it is given by 
%%%
\begin{equation}
m_{\gamma_D}
\simeq
60
\sqrt{1+z}
~
\,
\bigg[\frac{\alpha_D}{10^{-2}} \bigg]^{1/2} 
\bigg[\frac{m_{e_D}}{1 \, \textrm{MeV}} \bigg]^{1/4}
\bigg[\frac{f}{10^{-2}} \bigg]^{1/4} 
\bigg[\frac{\lambda_C}{1\, \textrm{Mpc}} \bigg]^{1/4} 
~ 
\mathrm{keV}
\quad .
\label{eq:photonmassconstraint}
\end{equation}

In order to obtain predictions for small-scale structure formation that depart from the CDM paradigm,
in this work we focus on the case in which our dark sector is collisional. 
For this reason, 
we commit to dark-photon masses below the threshold value in Eq.~\eqref{eq:photonmassconstraint} throughout matter domination, 
\textit{i.e.}, up to $z \simeq 0$.
For a perturbation of the size of the Milky-Way galaxy, $\lambda_C=4 \, \textrm{Mpc}$, and typical model parameter values $m_{e_D}=1~\textrm{MeV}$, $\alpha_D=1/100$,
dark-electrons remain kinetically coupled for $m_{\gamma_D} \leq 90 \, \textrm{keV}$.
This ensures, for instance, that as dark-electron halos form via gravitational collapse, they can heat up, as baryons do. 

Finally, from Fig.~\ref{fig:one}, we see that for small $m_{e_D}$, 
where Compton scattering is efficient (c.f.~Eq.~\eqref{eq:compton}),
the gas is in the tightly coupled regime.
In general,
for any perturbation with current (or comoving) size $\lambda_C=(1+z)\lambda_P$ there is a dark-electron mass below which the gas is tightly coupled.
Setting $\ell_{\gamma_D}=\lambda_P$ and using Eqns.~\eqref{eq:meanfreepaths} and \eqref{eq:compton},
this threshold value for $m_{e_D}$ is given by
%%%
\begin{equation}
m_{e_D} \simeq
~ 
2 \times  10^{-2}
\,
(1+z)^{2/3}
\,
\bigg[\frac{\alpha_D}{10^{-2}} \bigg]^{2/3} 
\bigg[\frac{f}{10^{-2}} \bigg]^{1/3} 
\bigg[\frac{\lambda_C}{1\, \textrm{Mpc}} \bigg]^{1/3} 
\, \textrm{MeV}
\quad .
\label{eq:meddecoupled}
\end{equation}
%%%

Equipped with the expressions for the speed of sound of dark electrons in each thermodynamic regime,
we are now ready to study the growth of dark-electron overdensities using linear perturbation theory.
We present the linear analysis in the following section.

\subsection{Dynamics of linear-perturbation growth}
\label{sec:lineardynamics}
In this section, we identify the conditions to ensure the growth of dark-electron perturbations in the linear regime.
A summary of the evolution and rate of growth of cold dark matter (CDM), 
interacting matter, and radiation overdensities as a function of the scale factor is given in Table \ref{tab:growthlinear}.
Both CDM and interacting-matter perturbations grow significantly only after matter-radiation equality. 
For the dark-electron perturbations to grow starting at some redshift after equality,
they must be non-relativistic and satisfy the Jeans condition to ensure that gravity overcomes the restoring dark-electron pressure force at that redshift and at later times. 
Since we are not requiring dark-photon overdensities to necessarily collapse, 
dark photons may be either relativistic or non-relativistic. 

%%%%%%%%%%%%%%
%%% BEGINNING OF TABLE %%
\begingroup
\setlength{\arraycolsep}{9pt}
 \renewcommand\arraystretch{1.5}
\begin{table}[h!]
\begin{center}
$
\begin{array}{ccc} 
&  \text{Radiation domination} & \text{Matter domination}
\\
\hline
% \multirow{5}{*}{Inside horizon}  
 \text{CDM} 
& \log(a) 
& a
\\
\cline{1-3}
  \text{Interacting Matter}  
& \begin{array}{cccc} \lambda_P & < & \lambda_J^{\text{m}} & \quad - \\ \lambda_P & > &  \lambda_J^{\text{m}}  & \quad  \log(a) \end{array}
&  \begin{array}{cccc} \lambda_P & < & \lambda_J^{\text{m}} & \quad - \\ \lambda_P & > &  \lambda_J^{\text{m}}  & \quad  a \end{array}
\\
\cline{1-3}
  \text{Radiation} 
  & -
  & -
\\
\cline{1-3}
\end{array}
$
\end{center} 
\caption{
Growth of sub-horizon linear perturbations of proper scale in the conformal Newtonian gauge as a function of the scale factor $a$.
The hyphen  ``$-$'' stands for no growth.
$\lambda_J^{\text{m}}$ stands for the Jeans length of radiation or of an interacting matter component.
All super-horizon-sized perturbations are constant in the conformal Newtonian gauge.
The growth of CDM, interacting matter, and radiation perturbations is studied in detail in~\cite{Dodelson:1282338,Longair:1339212,Gorbunov:2011zzc,Kolb:206230,Ma:1995ey,Mukhanov:991646,vdb}.
}
\label{tab:growthlinear}
\end{table}

%%% END OF TABLE %%
%%%%%%%%%%%%%%

According to the Jeans condition,
a dark-electron perturbation with proper scale $\lambda_P$ is unstable to gravitational collapse if the scale exceeds the proper Jeans length,
$\lambda_P \geq \lambda_J$.
The proper Jeans length is given by
%%%
\begin{equation}
\lambda_J
=
c_s
\bigg(
\frac{\pi}
{\rho G}
\bigg)
^{1/2}
\quad ,
\label{eq:Jeansmass}
\end{equation}
%%%
where $c_s$ is the dark-electron speed of sound, given by Eqns.~\eqref{eq:idealgascs} and~\eqref{eq:tightcs},
and $\rho$ is the total matter density,
which in this section corresponds to the background dark-matter density $\rho^{\textrm{DM}}_0$.
The speed of sound depends on the dark-electron and dark-photon temperatures.
The dark-electron temperature is a free parameter in our model.
The dark-photon temperature, 
on the other hand,
remains equal to the dark-electron temperature up to kinetic decoupling. 
For a relativistic dark photon,
the thermal-decoupling temperature may be estimated by comparing the rate of relativistic Compton-energy transfer to dark electrons with the Hubble expansion rate~\cite{Peebles:1994xt}
\begin{eqnarray}
\nonumber
H(z)
&=&
\frac{\sigma_C \rho_{\gamma_D}^0}
{m_{e_D}}
\\
&=&
\frac{\pi^2}{15}
\frac{\sigma_C}
{m_{e_D}}
T_{\gamma_D}^4
\quad ,
\label{eq:thermaldecoupling}
\end{eqnarray}
where in the second line we made use of the relativistic energy density $\rho_{\gamma_D}^0=(\pi^2/15) \, T_{\gamma_D}^4$ and $H(z)$ is the Hubble expansion parameter.
For a fixed relativistic dark-photon to Standard Model temperature ratio, one may obtain the redshift $z$ as a function of  $T_{\gamma_D}$,
so the relation \eqref{eq:thermaldecoupling} can 
be numerically solved to find the Standard Model and dark sector temperature $T_{e_D , \gamma_D}^{\textrm{kin.\,dec}}$ at decoupling of the dark-electron and dark-photon temperatures.

If $T_{e_D , \gamma_D}^{\textrm{kin.\,dec}}\lesssim m_{\gamma_D}$, 
the dark electrons and dark photons decouple while dark photons are already non relativistic. 
In this case, 
after kinetic decoupling both the dark-electron and dark-photon temperatures redshift as $a^{-2}$, 
so both sectors remain at the same temperature until today.\footnote{Note that Eq.~\eqref{eq:thermaldecoupling} is not valid for a non-relativistic dark-photon,
but it suffices to find the boundary where kinetic decoupling happens when dark-photons are already non-relativistic $T_{e_D , \gamma_D}^{\textrm{kin.\,dec}}= m_{\gamma_D}$.}
On the other hand, if $T_{e_D , \gamma_D}^{\textrm{kin.\,dec}}\gtrsim m_{\gamma_D}$,
the dark electrons and dark photons decouple while dark photons are relativistic,
so the dark-electron sector becomes comparatively colder than the dark-photon sector by a factor $a^{-1}$. 
Note however that regardless of the dark-photon mass,
Compton-energy transfer is efficient in the tightly coupled regime, 
so for the purpose of calculating the dark-electron speed of sound in the tightly coupled regime one may always set $T_{e_D}=T_{\gamma_D}$. 

%%%%%%
% FIGURE
%%%%%%
\begin{figure}[t!]
\begin{center}
\includegraphics[width=15cm]{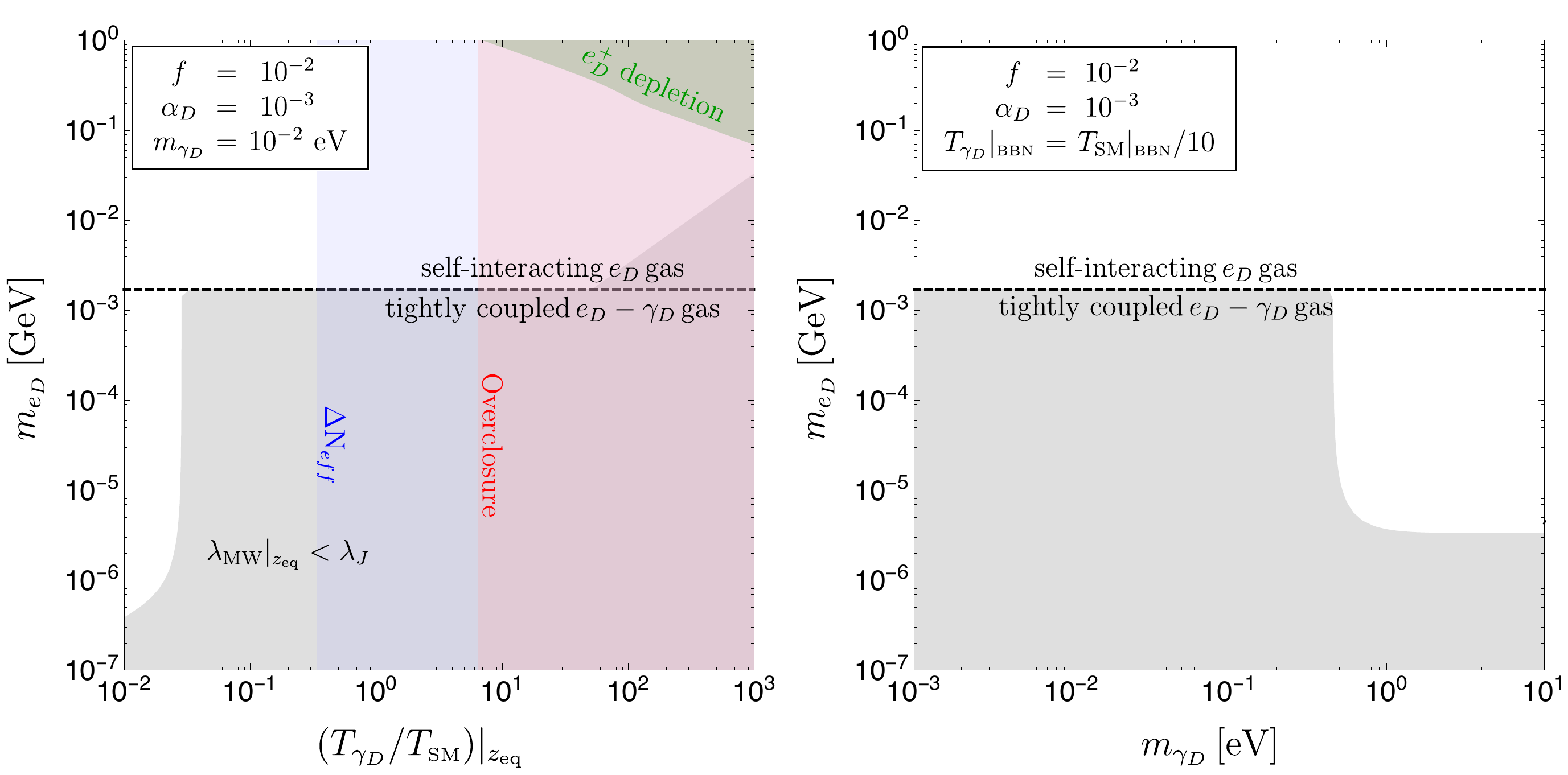}
\caption{
\textbf{Left:} 
In gray are 
regions of dark-electron mass and dark-photon to SM
temperature ratio that do \textit{not} satisfy the Jeans criterion at matter-radiation equality $z_{\textrm{eq}}=3400$ for linear growth of Milky-Way sized dark-electron perturbations,
$\lambda_{\textrm{MW}}|_{z_\textrm{eq}} \equiv \lambda_{\textrm{MW}}/(1+z_{\textrm{eq}}) > \lambda_J$, $\lambda_{\textrm{MW}} = 4\, \textrm{Mpc}$, for $f\equiv {\rho^{e_D}_0}/
{\rho^{\textrm{DM}}_0} = 1\%$, $\alpha_D=10^{-3}$, and fixed $m_{\gamma_D}= 0.01 \, \textrm{eV}$.  
For the chosen dark-photon mass,
dark photons are relativistic at equality in all the parameter space in the figure.
The dashed-black line indicates the transition between dark-electron thermodynamic regimes,
from a tightly-coupled gas of dark electron and dark photons, 
to an interacting gas of dark electrons that are decoupled from dark photons. 
The red and blue regions are excluded by dark-photon abundance overclosure and $\Delta N_{\textrm{eff}}$, 
respectively; see Eqns.~\eqref{eq:overclosure} and~\eqref{eq:neff}. 
The green region is not compatible with an effectively asymmetric dark sector due to inefficient dark-positron annihilations, Eq.~\eqref{eq:symmetricabundance}.
\textbf{Right:} Same as left, but as a function of the dark-electron and dark-photon masses,
for a fixed present-day dark-photon temperature $T_{\gamma_D}|_{\textrm{BBN}}=T_{\textrm{SM}}|_{\textrm{BBN}}/10$.
}
\label{fig:two}
\end{center}
\end{figure}
%%%%%%
% END OF FIGURE
%%%%%%

To demonstrate if linear growth of perturbations happens for the interacting dark-electron thermodynamic regimes,
we show in Fig.~\ref{fig:two} the regions of parameter space satisfying the Jeans criterion at matter-radiation equality for a perturbation on a comoving scale containing the Milky Way mass,
$\lambda_{\textrm{MW}}=4 \, \textrm{Mpc}$.
On the left panel of Fig.~\ref{fig:two}, we show such regions as a function of the dark-electron mass and dark-photon to SM
temperature ratio, while on the right panel we present the regions as a function of the dark-electron and dark-photon masses. 
The boundary of the gray region is set by $\lambda_{\textrm{MW}}/(1+z_{\textrm{eq}})=\lambda_J$.
Away from the tightly-coupled regime, 
in order to obtain this boundary we must specify the dark-electron temperature.
In this case, we have checked using Eq. \eqref{eq:thermaldecoupling} that along the boundary
the dark-electrons and dark-photons are kinetically coupled at equality, 
so we may set $T_{e_D}=T_{\gamma_D}$ for the purpose of obtaining this boundary.
We note that depending on $m_{e_D}$, 
the dark-electron gas may be in the tightly-coupled or self-interacting regimes at equality, 
as discussed in the previous section. 
The two regimes are separated in Fig.~\ref{fig:two} by the boundary Eq.~\eqref{eq:meddecoupled} in dashed black.
We also observe that galactic-sized dark-electron perturbations may grow after equality both in the self-interacting and tightly coupled regimes.
Note that this is in stark contrast with the baryonic sector, 
where the baryonic gas is tightly coupled with photons at equality, and grow of galactic-sized perturbations cannot happen before recombination.

Galactic-sized dark-electron perturbations can grow in the tightly-coupled regime, since dark photons may be non-relativistic and their abundance small,
so that dark electrons do not feel a large radiation pressure.
This can be seen in Fig.~\ref{fig:two} (left) at small temperatures,
or Fig.~\ref{fig:two} (right) at large dark-photon masses. 
From Fig.~\ref{fig:two} and from Eq.~\eqref{eq:meddecoupled}, 
we note that galactic-sized perturbations are in the tightly coupled regime throughout matter domination typically if $m_{e_D} \lesssim 1\, \textrm{MeV}$ (for the parameter choice shown in the figure).
In Section~\ref{sec:four}, we will see that the formation of compact objects from dark electron halo fragmentation typically happens for $m_{e_D} \gtrsim 1\, \textrm{MeV}$,
so in what follows we commit to this range for the dark-electron mass. 
For $m_{e_D} \geq 1\, \textrm{MeV}$, 
galactic-sized dark-electron perturbations decouple from the dark photons during or before matter domination for all the parameter space considered in this work,
so we need to ensure that linear perturbations may grow in the self-interacting gas regime. 

In the self-interacting gas regime, 
perturbations may grow linearly if they overcome
the dark-electron pressure Eq.~\eqref{eq:pressure}.
At the low densities typical of the linear regime and for large enough dark-photon masses we may take the dark-electron pressure to be mostly kinetic, $P_{e_D} \simeq n_{e_D}T_{e_D}$.\footnote{The pressure
due to the repulsive dark-photon force will be important later on deep in the non-linear regime,
where the dark-electron densities are high.}
As a consequence,
the Jeans criterion is satisfied as long 
as the dark-electron sector is cold enough.
In fact, 
imposing that overdensities of the size of our galaxy may grow, $\lambda_{\textrm{MW}}/(1+z) > \lambda_J$, leads to a constraint on the dark electron temperature given by 
\begin{equation}
T_{e_D} < 80 \, \bigg[\frac{m_{e_D}}{1 \, \textrm{MeV}}\bigg] \, \mathrm{K} \simeq 30 \, \bigg[\frac{m_{e_D}}{1 \, \textrm{MeV}}\bigg]  \, T_{\textrm{SM}}  \quad .
\label{eq:temperaturecollapse}
\end{equation}
Note that the constraint Eq.~\eqref{eq:temperaturecollapse} is redshift independent, 
since both the Jeans and Milky-Way dimensions grow linearly with the scale factor.
It turns out that the condition Eq.~\eqref{eq:temperaturecollapse} on the dark-electron temperature is automatically satisfied once we impose the limits Eq.~\eqref{eq:overclosure} and Eq.~\eqref{eq:neff} from dark-photon overclosure and from the number of relativistic species $N_{\textrm{eff}}$.
The conditions~\eqref{eq:overclosure} and \eqref{eq:neff} are limits on the dark-photon temperature at dark-photon chemical decoupling and BBN,
but they can be translated into limits in the dark-electron temperature at matter domination as follows.
First, 
if dark electrons and dark photons are in thermal contact down to matter domination and if dark-photons remain relativistic,
the dark electron temperature redshifts as $1/a$ down to matter domination,
and the limits Eqns.~\eqref{eq:overclosure} and \eqref{eq:neff} apply directly on $T_{e_D}$ at matter domination (up to factors of $g_{*S}^{1/3}$).
In this case we immediately see that the limits Eqns.~\eqref{eq:overclosure} and \eqref{eq:neff} are an order of magnitude more stringent than the condition \eqref{eq:temperaturecollapse}.
This is the case shown in Fig.~\ref{fig:two}, 
where the limits from $N_{\textrm{eff}}$ and overclosure are indicated by blue and red contours, respectively. 
On the other hand,
in the scenario where the dark electron temperature redshifts non-relativistically (as $1/a^2$) before or during matter domination, 
the dark electron to SM temperature ratio becomes much smaller than the ratios in Eqns.~\eqref{eq:overclosure} and \eqref{eq:neff},
so the condition \eqref{eq:temperaturecollapse} is trivially satisfied.

While in Fig.~\ref{fig:two} we presented an analysis of the Jeans condition at matter-radiation equality,
ensuring growth of perturbations at equality is a sufficient but not necessary condition for the dark-electron perturbations to efficiently follow the dominant CDM component into the non-linear regime.
In practice,
it suffices to satisfy the Jeans criterion a few Hubble times before CDM peturbations become non-linear. 
The reason is that after dark-electron perturbations start growing,
their difference with the dominant CDM dark-matter perturbations $(\delta_{\textrm{CDM}}-\delta_{e_D})/\delta_{\textrm{CDM}}$ decreases with the scale factor approximately as $1/a$~\cite{Gorbunov:2011zzc},
so dark-electron perturbations quickly catch up with the CDM perturbations.
If the temperature of our dark sector is below the limit in Eqns.~\eqref{eq:overclosure} and.~\eqref{eq:neff},
our dark-electron perturbations are guaranteed to grow after they transition from the tightly coupled into the self-interacting regimes. 
For all the parameter space considered in the rest of this work, $10^{-3} \leq \alpha_D \leq 10^{-1}$, $m_{e_D} \geq 1 \, \textrm{MeV}$, $f \leq 10^{-1}$,
we have checked that such transition happens early enough.
In addition,
note that once the Jeans criterion is satisfied at some redshift for a perturbation of physical scale $\lambda_P$ it is also satisfied at later times, 
so the dark-electron perturbations are guaranteed to become non-linear at some point.
The reason is that from Eq.~\eqref{eq:idealgascs} we see that in the dark electron interacting gas regime,
the dark electron speed of sound decreases at least as $\sim a^{-1/2}$. 
As a consequence, the Jeans length in Eq.~\eqref{eq:Jeansmass}, 
increases at most as $\sim a$, 
which is the same growth rate as the scale of the perturbation $\lambda_P$. 
We estimate the redshift at which galactic-sized dark electron perturbation become non-linear in the next section.

Based on the previous discussion, we choose a dark sector with the following parameters:  
\begin{enumerate}
\item dark-photon mass below the threshold value in Eq.~\eqref{eq:photonmassconstraint} setting $\lambda_C=4 \, \textrm{Mpc}$ and $z=0$,
\begin{equation}
m_{\gamma_D}
\leq
85
~
\,
\bigg[\frac{m_{e_D}}{1 \, \textrm{MeV}} \bigg]^{1/4}
\bigg[\frac{\alpha_D}{10^{-2}} \bigg]^{1/2} 
\bigg[\frac{f}{10^{-2}} \bigg]^{1/4} 
~ 
\mathrm{keV}
\quad ,
\label{eq:mgammacondition}
\end{equation}
\item fine-structure constant $10^{-3} \leq \alpha_D \leq 10^{-1}$ ,
\item dark-electron mass $m_{e_D} \geq 1 \, \textrm{MeV}$, 
\item dark-sector temperature satisfying  the limits Eqns.~{\eqref{eq:symmetricabundance}},~\eqref{eq:overclosure} and~\eqref{eq:neff}, 
\item dark-electron to dark-matter-background density ratio $f \leq 10\%$. 
\end{enumerate}
For these parameters, the dark electrons on galactic scales are a self-interacting gas in kinetic equilibrium throughout matter domination, 
and the corresponding galactic-sized perturbations are guaranteed to grow gravitationally until they become non-linear. 
The first and second conditions in the item list above ensure dark electrons are collisional throughout matter domination so they may heat up after halo collapse. 
The second, third, and fourth conditions ensure that dark electrons decouple from dark photons during or before matter domination and track the CDM perturbations into the non-linear regime, 
that the dark electron sector is effectively asymmetric (dark positrons efficiently annihilated in the early Universe), 
and that bounds from overclosure and nucleosynthesis are avoided. 
The last condition ensures that bounds from self-interactions in the dark sector are evaded.
The conditions above are sufficient, but not necessary for dark electron halos to transition into the non-linear regime and to efficiently heat up after collapse. 
There are other regions of parameter space leading to an interesting structure formation history for our dark sector, 
but for concreteness and brevity we commit to the above choices and we now move on to discuss the transition into the non-linear regime.

\subsection{Turnaround and transition into the non-linear regime}
\label{sec:transition}
In linear theory the perturbations are coupled to the expanding Hubble background. 
As the overdensities grow, 
they eventually become non-linear self-gravitating bodies and ``turn around'',
so they stop expanding with the Hubble flow. 
The typical density at which perturbations turn around may be calculated using the spherical collapse model~\cite{2010gfe..book.....M}. 
The turnaround overdensity in the spherical collapse model is 
%%%
\begin{equation}
\delta(z_{\textrm{ta}})
=
\frac{9\pi^2}{16}
\quad ,
\end{equation}
%%%
where $z_{\textrm{ta}}$ is the redshift at turnaround.
The corresponding dark-electron gas density at turnaround is then given by
\begin{eqnarray}
\nonumber 
\rho_{e_D}
&\simeq&
\frac{9\pi^2}{16} \rho^{e_D}_0
=
\frac{9\pi^2}{16} f \rho^{\textrm{DM}}_0
\label{eq:ics}
\end{eqnarray}
where $f$ is the dark electron fraction of the dark matter density (c.f. Eq.~\eqref{eq:DMfraction}).

On the other hand,
the turnaround redshift roughly corresponds to the time at which the perturbations become non-linear, 
$\delta \simeq 1$.
To determine the redshift at which perturbations become non-linear, 
we must specify the initial conditions for our dark-electron perturbations.
They are set by the primordial power spectrum 
%%%
\begin{equation}
\big<
\delta_{\mathbf{k}} \delta_{\mathbf{k}}^*
\big>
\equiv
(2\pi)^3
P(k)
\,
\delta^3(0)
\quad ,
\label{eq:powerspeccontdef}
\end{equation}
%%%
which we take to be the usual Harrison-Zeldovich initial spectrum normalized to the CMB perturbations with $\sigma_8=0.83$~\cite{Ade:2015xua}. 
Dark-electron perturbations on scales larger than the Jeans length grow throughout matter domination, 
and their power spectrum evolves roughly as the cold or warm dark matter power spectrum. 
For concreteness, consider the cold dark matter power spectrum.
The power spectrum in the linear regime during matter domination as a function of redshift is given by~\cite{2010gfe..book.....M,1986ApJ...304...15B}
%%%
\begin{equation}
P(k,z)=
\frac{1}
{(1+z)^2}
T(k)^2
P(k)
\quad ,
\end{equation}
%%%
where $T(k)$ is the transfer function, which is given by
%%%
\begin{equation}
T(k)
=
\frac{\log(1+2.34 q)}
{2.34 q}
\big[
1
+
3.89 q
+
(16.1 q)^2 
+
(5.46 q)^3
+
(6.71 q)^4
\big]^{-1/4} 
\end{equation}
%%%
and $q=k/(\Omega_0 h^2 \mathrm{Mpc}^{-1})$. 
The non-linear regime starts at a redshift $z_{\textrm{ta}}$ when the dark-electron perturbations are roughly equal to one,
%%%
\begin{equation}
\frac{k^3}
{2\pi^2}
P(k,z_{\textrm{ta}})
=
1
\quad .
\label{eq:dimlesspw}
\end{equation}
%%%
The solution of Eq~\eqref{eq:dimlesspw} for a perturbation on the scale of the Milky-Way Lagrangian radius, $k=(2\pi/4\, \textrm{Mpc})$ is $z_{\textrm{ta}} \simeq 1.5$.  
This is of course nothing else than a rough estimate of the age of the assembly of our galaxy. 
The dependence of $z_{\textrm{ta}}$ on the size of the perturbation is only logarithmic, 
so non-linearities in the dark-electron sector on a wide range of scales arise approximately at similar redshift. 

\FloatBarrier
\section{Non-linear collapse of a dark-electron galactic halo}
\label{sec:four}
While the analysis of linear perturbations in the dark-electron sector is rather straightforward,
the subsequent non-linear evolution of the overdensities is much more complicated,
and a detailed study of non-linearities requires heavy numerical simulations.
Nevertheless,
in this section we show that many of the features of the non-linear evolution of the dark-electron perturbations may be understood using energy conservation and halo-stability arguments.
For concreteness we focus on the evolution of dark electrons on galactic scales and smaller,
but our analysis can be easily extended to study dark-electron structure on larger scales. 

We proceed as follows.
In Section~\ref{sec:discussion}, we discuss generalities on the conditions for halo collapse, fragmentation, and the complications of the analysis. 
In Section~\ref{sec:contourequation}, 
we present a simple equation that describes the non-linear dark-electron halo evolution.
In Section~\ref{sec:stages}, 
we study the dark-electron halo collapse.
In Section~\ref{sec:smallest}, we calculate the mass of the smallest fragments within the dark-electron halo, 
which ultimately lead to the formation of exotic compact objects.
Finally, in Section~\ref{sec:angularmomentum}, we discuss the effects of including angular momentum for the dark-electron halo.

\subsection{Preliminaries: halo fragmentation from a Jeans analysis}
\label{sec:discussion}
After turnaround, 
the cold dark matter, 
baryonic and dark-electron overdensities continue to collapse.
As the density and pressure start to increase, 
subsequent collapse of a dark-electron overdensity happens only if the mass of the overdensity $M$ exceeds the Jeans mass
\begin{equation}
M \geq 
m_J  
\quad ,
\quad
m_J  
\equiv
\frac{4\pi}
{3}
\bigg(\frac{\lambda_J}{2}\bigg)^3
\,
\rho_{e_D} 
=
\frac{\pi}{6}
c_s^3
\bigg(
\frac{\pi}
{\rho G}
\bigg)
^{3/2}
\rho_{e_D} ,
\label{eq:Jeanscrit}
\end{equation}
where $\rho$ is the sum of the cold dark matter, baryonic and dark-electron mass densities,
which determine the gravitational potential,
and $\rho_{e_D}$ is just the dark-electron mass density within the halo.
Importantly, 
the dark-electron Jeans mass may be much smaller than the dark-electron halo mass $M_{\textrm{halo}}^{e_D}$.
In this case, as the halo collapses,
smaller sub-halos start to collapse on their own and form substructure. 
This dynamical process, 
which is called \textit{fragmentation},
is the origin of substructure in galaxies for both baryonic and dark-electron perturbations~\cite{1953ApJ...118..513H}.
Note that the smallest fragment that can collapse at some point in the halo evolution has a mass $m_{J}$ and volume $V_J \sim \lambda_J^3$.

As the collapse proceeds, the Jeans mass and therefore the size of the smallest fragment changes, 
since the Jeans mass depends on the dark-electron density and temperature. 
Depending on how the Jeans mass evolves, 
this may either trigger or stop fragmentation,
and lead to larger or smaller fragments.
The point of halo evolution at which the Jeans mass reaches its minimal value is called the \textit{end of fragmentation},
and at that point the halo may divide into \textit{minimal fragments}.
These minimal fragments in the baryonic sector  are the protostars that are the seeds of our galaxy's stars. 
Equivalently,
the minimal fragments in the dark-electron sector give rise to exotic compact objects.

A detailed study of the evolution of the dark-electron halo and of the process of fragmentation
is very intricate for at least two important reasons. 
First, 
it requires studying not only the dark-electron component, 
but also the evolution of baryons and the rest of the dark matter,
since the gravitational potential into which dark-electrons collapse and the Jeans mass in Eq.~\eqref{eq:Jeanscrit} are determined by the sum of all matter components, or at least by the dominant one.
As discussed in the previous section, during the linear regime all matter perturbations grow adiabatically with the scale factor, 
so the matter density in a galactic protohalo at the onset of the non-linear regime is dominated by the main cold (or warm) dark-matter component.
However, 
while dark-electron and baryonic overdensities initially fall  into the galactic gravitational potential set by this primary cold dark matter component,
they further collapse into the center of the halo, 
since their gravitational energy is converted into thermal energy, 
and since thermal energy and pressure support can be released through cooling.
Once the dark-electron and baryonic components exceed the cold dark matter density in the center of the halo, 
they continue to collapse under their own gravity and cold dark matter can be neglected~\cite{vdb}.

On the other hand, 
it is likely that the evolution of baryonic and dark-electron overdensities within the galaxy is correlated.
Nevertheless, and for simplicity,
in what follows we assume that neglecting the baryonic gravitational potential does not lead to significant differences in our conclusions.
However, we note that including the baryonic gravitational potential can only help with the formation of dark-electron substructure since it lowers the dark-electron Jeans mass.
In addition, in some regions of the galaxy baryons may dominate, especially near the center of halos and along galactic disks,
but dark electrons may dominate and form substructure away from these regions.
In particular, 
even within a disk galaxy, it may happen that the dark-electron halo does not form a disk~\cite{1991ApJ...380..320N,Ghalsasi:2017jna}, 
or forms a disk that is on a different orientation than the baryonic disk~\cite{Fan:2013tia},
or that dark electrons form structure before falling into a disk. 

In summary, neglecting baryons and CDM, the Jeans mass in Eq.~\eqref{eq:Jeanscrit} simplifies to 
%%%
\begin{equation}
m_J  
=
\frac{\pi}{6}
c_s^3
\,
\bigg(
\frac{\pi}
{G}
\bigg)
^{3/2}
\bigg(
\frac{1}
{\rho_{e_D}}
\bigg)
^{1/2}
\quad ,
\label{eq:Jeanscrit2}
\end{equation}
%%%
where the dark-electron speed of sound is given in Eq.~\eqref{eq:idealgascs}.
This simplification implies that we can study the late-time evolution of the dark-electron Jeans mass solely by analyzing the evolution of the dark-electron density and temperatures.
 
The second complication in the analysis of the evolution of the dark-electron component,
is that the collapse of the dark-electron halo is a highly inhomogeneous process.
Even if the halo starts as a roughly homogeneous gas,
accumulation of matter near the center and fragmentation eventually lead to large inhomogeneities in temperature and density. 
We ignore this important complication, 
and analyze the collapse assuming homogeneity throughout, at least in parts of the halo.
Even if this assumption is extremely simplistic,
in the next section we will see that it allows us to understand in detail the process of dark-electron halo fragmentation.

\subsection{Temperature- and density-evolution equations for the dark-electron clumps}
\label{sec:contourequation}
In this section, we model the evolution of the density, temperature, and size of the smallest dark-electron fragments as the dark-electron halo collapses. 
The analysis we present was first carried out for the baryonic sector in the seminal works~\cite{doi:10.1093/mnras/176.2.367,1977MNRAS.179..541R,1977ApJ...211..638S}.

The dark-electron Milky-Way halo may be divided into approximately homogeneous and roughly independent regions,
which we refer to as dark-electron clumps,
over which the density and temperature are constant in space (so the Jeans mass in these regions is also constant).
The maximal volume of such regions is denoted by $V \leq V_{\textrm{MW}}$, 
and contains the mass $M \leq M_{\textrm{MW}}^{e_D}$.
In the initial stages of collapse the whole dark-electron halo is roughly homogeneous, and the clump just corresponds to the  whole dark-electron halo, $V \sim V_{\textrm{MW}}$, $M \sim M_{\textrm{MW}}^{e_D}$.
The mass of the dark-electron halo may be estimated from the mass of the host galaxy and the dark-electron to dark-matter ratio as
\begin{equation}
M_{\textrm{halo}}^{e_D}= f M_{\textrm{halo}} \quad  .
\label{eq:halomass}
\end{equation}
As the halo collapses, starts to fragment, and becomes inhomogeneous,
the largest clumps that can be treated as roughly homogeneous are just the smallest fragments of the halo, 
with volume $\sim \lambda_J^3$.

The evolution of the density and temperature within the clump may be obtained using energy conservation.
The thermal energy per unit mass of the homogeneous gas enclosed in the volume $V$ is
\begin{equation}
e_{\textrm{kin}}
=
\frac{3 T_{e_D}}{2 m_{e_D}}  \quad .
\end{equation}
Collapse of the dark-electron gas leads to an increase of the thermal energy
while cooling releases energy, 
so the time evolution of the dark-electron temperature (using the first law of thermodynamics) is described by
\begin{equation}
\frac{d e_{\textrm{kin}} }
{dt}
=
\frac{3}{2 m_{e_D}} \frac{d T_{e_D}}{dt}
=
-\frac{P_{e_D}}{M}
\frac{dV}{dt}
-\Lambda
\quad ,
\label{eq:intstep1}
\end{equation}
where $P_{e_D}$ is the dark-electron pressure Eq.~\eqref{eq:pressure}, 
the first term on the right-hand side accounts for the work done by the collapsing gas and $\Lambda$ is the cooling rate per unit mass.
The change in the gas volume due to the collapse may be related to the change in the dark-electron gas mass density
\begin{equation}
\frac{dV}{dt}
=
M
\frac{d}{dt}
\bigg(
\frac{1}{\rho_{e_D}}
\bigg)
\quad .
\label{eq:volumechange}
\end{equation}
Using Eq.~\eqref{eq:volumechange} in Eq.~\eqref{eq:intstep1} we get
\begin{equation}
\frac{3}{2 m_{e_D}} \frac{dT_{e_D}}{dt}
=
\frac{P_{e_D}}{\rho_{e_D}^2}
\frac{d\rho_{e_D}}{dt} 
-\Lambda
\quad .
\label{eq:contour1}
\end{equation}
Eq.~\eqref{eq:contour1} describes the evolution of the dark-electron temperature and mass density as the gas collapses.
This expression also allows us to find a simple relation between the temperature and density as collapse proceeds.
The relation is obtained by rewriting Eq.~\eqref{eq:contour1} as
\begin{equation}
\frac{d \log T_{e_D}}
{d \log \rho_{e_D}}
=
\frac{2 }{3} \frac{m_{e_D} P_{e_D}}{\rho_{e_D} T_{e_D}}
-2 
\,
\frac{ t_{\textrm{collapse}}}{t_{\textrm{cooling}}}
\quad ,
\label{eq:contour2}
\end{equation}
where we defined collapse and cooling times 
\begin{equation}
 t_{\textrm{collapse}}
\equiv
 \bigg(\frac{d \log \rho_{e_D}}{dt} \bigg)^{-1}
 \quad , \quad 
t_{\textrm{cooling}}
\equiv
\frac{3 T_{e_D}}{m} \frac{1}{\Lambda} 
\quad .
 \label{eq:tdefinition}
\end{equation}
To determine the evolution of the gas temperature and density using Eq.~\eqref{eq:contour2}, 
we must specify the typical collapse and cooling times. 
We start by discussing the collapse time.

For a pressureless spherical gas clump, 
the collapse time is simply the time that it takes the gas to free-fall into the center of the perturbation.
The free-fall time is~\cite{Penston:1969yy} 
\begin{equation}
t_{\textrm{ff}}
\equiv
\bigg(
\frac{1}
{16 \pi G \rho_{e_D}}
\bigg)^{1/2}  \quad .
\label{eq:freefalltime}
\end{equation}
Anytime the mass of the clump exceeds the Jeans mass, gravity overcomes pressure,
so we simply neglect the small pressure resistance and replace the collapse time by the free-fall time.
Since including the effects of pressure can only slow down collapse,
the free-fall time also gives an absolute lower limit on the typical collapse time.

In an initial stage of collapse, 
the gas has a low density and temperature so it mostly free-falls.
However, as it collapses the kinetic pressure rises and the dark-electron clumps are eventually stabilized once the Jeans mass becomes similar to the clump's mass.
At this stage, 
collapse close to the clump's virialized state may still proceed,
since kinetic energy can be slowly released due to cooling.
It is then clear that close to virialization,
the collapse time is of the order of the cooling time.
It is quite simple to confirm this expectation as follows.
Since the clump stays nearly virialized,
the collapse time must accommodate so that the contour specified by Eq.~\eqref{eq:contour2} matches the contour defined by the virial condition $M =m_J$. 
Pressure can be released via cooling if the main source of pressure is just kinetic, 
in which case the speed of sound entering into the Jeans mass expression Eq.~\eqref{eq:Jeanscrit2} is $c_s=\sqrt{T_{e_D}/m_{e_D}}$.
It is then clear that the contour of constant Jeans mass  $m_J=M$ leads to
\begin{equation}
\frac{d \log T_{e_D}}
{d \log \rho_{e_D}}
=
\frac{1}{3} \quad , 
\label{eq:contour3}
\end{equation}
in agreement with~\cite{1977MNRAS.179..541R}. 
By comparing Eq.~\eqref{eq:contour3} with Eq.~\eqref{eq:contour2} and setting the pressure to be the ideal gas pressure, 
we conclude that for nearly virialized collapse of an ideal dark-electron gas, 
the collapse time is equal to 
 $1/6 ~  t_{\textrm{cooling}}$.
Note that if the cooling time is shorter than the free-fall time,
the collapse time must of course  again be replaced by the free-fall time.
 Summarizing, 
the collapse timescale of our homogeneous dark-electron gas is
%%%
\begin{eqnarray}
\nonumber
&
 t_{\textrm{collapse}}
\equiv
\bigg(
\cfrac{d \log \rho_{e_D}}{dt}\bigg) ^{-1}
=
&
%%%
\\
%%%
&
\left\{
%%%%%%%%%
\begin{array}{ccc}
t_\textrm{ff}
\equiv
\bigg(
\cfrac{1}
{16 \pi G \rho_{e_D}}
\bigg)^{1/2}
&
M > m_J 
&
\textrm{Adiabatic free-fall}
%%%
\\
%%%
\frac{1}{6} t_{\textrm{cooling}}
&
M = m_J
~
\textrm{and} \quad 
t_{\textrm{cooling}} > t_{\textrm{ff}} 
&
\textrm{Nearly virialized contraction} 
%%%%
%\\
%%%%
%t_\textrm{ff}
%&
%t_{\textrm{cooling}} \gtrsim t_{\textrm{ff}}
%&
%\quad
%\textrm{Collapse with fragmentation}
\end{array}
%%%%%%%%%
\right.
&
\nonumber 
\\
&&
 \label{eq:tdefinition2}
\end{eqnarray}
%%%

We now turn to discussing the cooling time. 
In our dark-electron model, 
cooling proceeds via  Brems\-strahlung
and Compton scattering off the background dark photons, 
as shown in Fig.~\ref{fig:cooling}.
Energy transfer to or from dark photons due to Compton scattering is efficient at large redshift
where the number density of background dark photons is large, 
but it is highly inefficient in the low redshift regime relevant for non-linear collapse, and so we neglect it. 
%%%%%%%%%%
% BEGIN FIGURE
%%%%%%%%%%
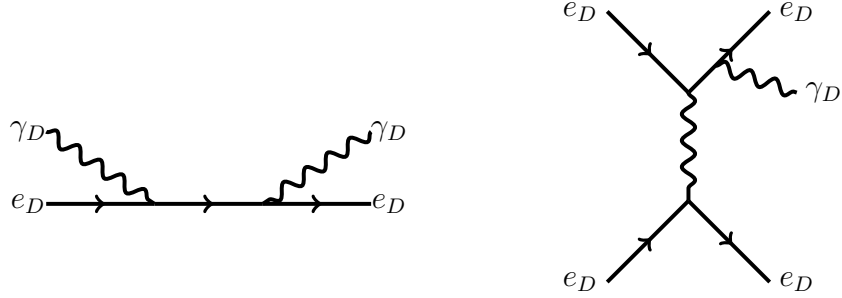
\begin{figure} [t!]
\begin{center}
\begin{tikzpicture}[line width=1.5 pt, scale=1.2,baseline=-1.3cm]
  \draw[vector]       (-1.2,0.8)--(0, 0.0);
  \draw[fermionbar]   (0, 0.0)-- (-1.2, 0);
  \draw[fermionbar]        (1.2, 0.0)--(0, 0.0);
  \draw[vector]       (2.4,0.8)--(1.2, 0.0);
    \draw[fermionbar]    (2.4, 0.0)--(1.2, 0);
\node at (-1.4, 0) {$e_D$};
\node at (-1.4, 0.8) {$\gamma_D$};
\node at (2.6, 0) {$e_D$};
\node at (2.6, 0.8) {$\gamma_D$};
\end{tikzpicture}
%%%%%%
\quad \quad \quad\quad
%%%%%%
\begin{tikzpicture}[line width=1.5 pt, scale=1.2]
  \draw[fermionbar]   (0, 0.0)-- (-0.9, 0.9);
    \draw[vector]       (0,0)--(0, -1.2);
  \draw[fermionbar]        (0,-1.2)--(-0.9, -2.1);
    \draw[fermionbar]    (0.9, 0.9)--(0, 0.0);
      \draw[vector]       (0.3,0.3)--(1.2, 0);
     \draw[fermionbar]        (0.9, -2.1)--(0,-1.2);
\node at (-1.2, 0.9) {$e_D$};
\node at (-1.2, -2.1) {$e_D$};
\node at (1.2, 0.9) {$e_D$};
\node at (1.2, -2.1) {$e_D$};
\node at (1.5, 0) {$\gamma_D$};
\end{tikzpicture}
\end{center}
\vspace*{-0.5cm}
\caption{Dark electron may cool only via scattering on dark CMB photons (Compton) or Brems\-strahlung. Compton energy transfer is inefficient at low redshift. Brems\-strahlung cooling is efficient at high dark-electron densities.}
\label{fig:cooling}
\end{figure}
%%%%%%%%%%
% END FIGURE
%%%%%%%%%%
On the other hand, Brems\-strahlung cooling is efficient at the large dark-electron densities typical of the non-linear regime. 
The Brems\-strahlung energy emission rate per unit dark-electron mass is~\cite{Haug:1975bk2}
\footnote{We add an exponential factor $e^{-m_{\gamma_D}/T_{e_D}}$ with respect to reference~\cite{Haug:1975bk2} to account for suppression of the Brems\-strahlung rate for $m_{\gamma_D} \geq T_{e_D}$.
Note that since the dipole moment of two electrons vanishes,
Brems\-strahlung emission is mostly quadrupole.
}
\begin{equation}
\Lambda_{\textrm{BS}}
=
\frac{32 \alpha_D^3 \rho_{e_D} T_{e_D}}{\sqrt{\pi}m_{e_D}^4} \sqrt{\frac{T_{e_D}}{m_{e_D}}}
e^{-m_{\gamma_D}/T_{e_D}}
\quad .
\label{eq:BS}
\end{equation}
Energetic dark photons may also be reabsorbed in the dark-electron gas via inverse Brems\-strahlung, 
as shown in Fig.~\ref{fig:absorption}. 
If the dark-photon absorption mean free 
path\footnote{Not to be confused with the dark-photon elastic scattering mean free path, given by Eq.~\eqref{eq:meanfreepaths}.} 
is larger than the size of the dark-electron region under consideration $\ell_{\gamma_D}^{\textrm{abs}} \gg V^{1/3}$, 
the gas is optically thin and dark photons may efficiently evacuate energy from the bulk of the gas volume.
In the opposite case $\ell_{\gamma_D}^{\textrm{abs}} \ll V^{1/3}$ the dark-electron clump is optically thick, and it may only cool from its surface,
so the cooling rate is suppressed.
The absorption mean free path for relativistic dark photons in a non-relativistic dark-electron gas is calculated in Appendix~\ref{app:two}.
It is given by
\begin{equation}
\ell_{\gamma_D}^{\textrm{abs}}
=
8.3 \times 10^{-3}
~
\frac{({m_{e_D}^{9} T_{e_D}^5})^{1/2}}
{\rho_{e_D}^2 \alpha_D^3}
\quad .
\label{eq:libs}
\end{equation}
To account for photon reabsorption,
we simply suppress the Brems\-strahlung bulk cooling rate by an exponential factor if the material is optically thick.
The dark-electron gas cooling rate per unit mass is then given by 
\begin{equation}
\Lambda=(\Lambda_{\textrm{Compton}} + \Lambda_{\textrm{BS}}) \, e^{-{V^{1/3} \sqrt{N_{\textrm{sc}}}}/{\ell_{\gamma_D}^{\textrm{abs}}}}
\quad \quad , 
\label{eq:coolingrate}
\end{equation}
($\Lambda_{\textrm{Compton}}\ll \Lambda_{\textrm{BS}}$) where the factor $\sqrt{N_{\textrm{sc}}} \geq 1$ in the optical-thickness exponential factor corresponds to the square-root of the number of dark-photon rescatterings,
and is introduced to account for the effective dark-photon displacement needed to escape the gas cloud as it undergoes a random walk due to elastic Compton scattering.
The number of scatterings is given by the squared ratio of the clump size to the Compton mean-free path~\cite{1986rpa..book.....R}
\begin{equation}
N_{\textrm{sc}}
=
\bigg(
\frac{V^{1/3}}{\ell_{\gamma_D}^{\textrm{C}}}
\bigg)^2
\quad ,
\end{equation}
where the Compton mean free-path is given by Eq.~\eqref{eq:meanfreepaths}. 

%%%%%%%%%%
% BEGIN FIGURE
%%%%%%%%%%
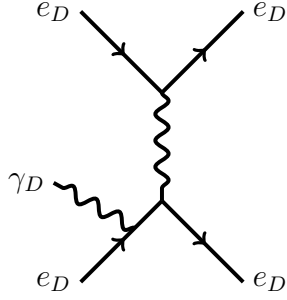
\begin{figure} [t!]
\begin{center}
%%%%%%
\begin{tikzpicture}[line width=1.5 pt, scale=1.2]
  \draw[fermionbar]   (0, 0.0)-- (-0.9, 0.9);
    \draw[vector]       (0,0)--(0, -1.2);
  \draw[fermionbar]        (0,-1.2)--(-0.9, -2.1);
    \draw[fermionbar]    (0.9, 0.9)--(0, 0.0);
      \draw[vector]       (-0.3,-1.5)--(-1.2, -1);
     \draw[fermionbar]        (0.9, -2.1)--(0,-1.2);
\node at (-1.2, 0.9) {$e_D$};
\node at (-1.2, -2.1) {$e_D$};
\node at (1.2, 0.9) {$e_D$};
\node at (1.2, -2.1) {$e_D$};
\node at (-1.5, -1) {$\gamma_D$};
\end{tikzpicture}
\end{center}
\vspace*{-0.5cm}
\caption{Inverse Bremsstrahlung dark-photon absorption.}
\label{fig:absorption}
\end{figure}
%%%%%%%%%%
% END FIGURE
%%%%%%%%%%

Finally, note that up to now we have not taken into account the effects from radiation pressure that arise whenever $N_{\textrm{sc}}\geq 1$, \textit{i.e.}, if the Compton mean free path is smaller than the size of the clump, $\ell_{\gamma_D}^{\textrm{C}} \leq V^{1/3}$.
Radiation pressure is in principle important whenever the Bremsstrahlung luminosity exceeds the Eddington luminosity,
\begin{equation}
\Lambda_{\textrm{BS}} \geq \frac{L_{\textrm{edd}}}{M} 
=
\frac{4\pi G m_{e_D}}{\sigma_C} \quad.
\label{eq:eddingtonlimit}
\end{equation}
If the clump luminosity is above the Eddington limit, 
the dark-electron gas cloud may be stabilized by radiation pressure.
However, 
radiation pressure may also be inefficient in stabilizing the clump, depending on how ``porous" is the gas~\cite{vdb,Owocki:2004zz,2015arXiv151103457K}.
In spatially inhomogeneous media, 
radiation selects regions of low density (``pores") to escape the cloud,
without exerting homogeneous pressure on the whole clump.
To account for these uncertainties we present two extremal cases:
for the remainder of the body of this paper
we study the halo evolution without the inclusion of radiation pressure,
while in Appendix \ref{app:C} we assume that radiation pressure may stabilize the dark-electron clumps and present the corresponding results.
In the end, 
we find that the typical size and mass of the exotic compact objects formed from fragmentation of the dark-electron halo are similar either
with or without the inclusion of radiation pressure.

\subsection{Stages of halo collapse and the end of fragmentation}
\label{sec:stages}

%%%%%%%%%%
% BEGIN FIGURE
%%%%%%%%%%
\begin{figure} [!]
\begin{center}
\includegraphics[width=15cm]{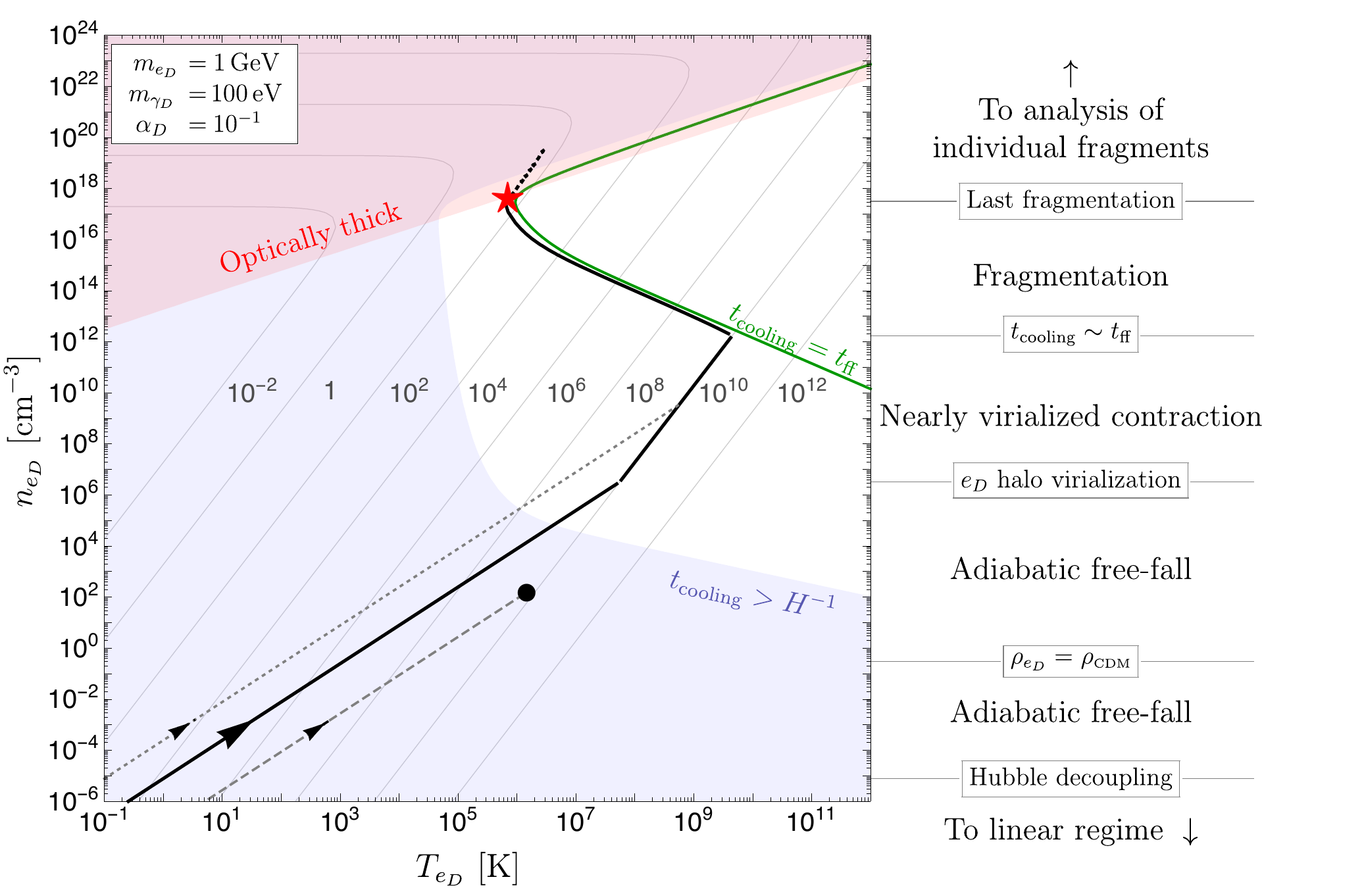}
\caption{\textbf{Solid black:} Temperature-density trajectory of a region of dark-electron gas as it collapses embedded in the 
dark-electron Milky-Way halo with mass $M_{\textrm{MW}}^{e_D}=10^{10} \, M_{\odot}$ (corresponding to dark-electron fraction $f=1\%$, c.f.~Eq.~\eqref{eq:halomass}),
for $m_{e_D}=1 \, \textrm{GeV}, m_{\gamma_D}=100 \, \textrm{eV}, \alpha_D=10^{-1}$.
The initial condition for the temperature of the collapse trajectory close to the turnaround density Eq.~\eqref{eq:ics} has been set to $T_{e_D} = 5 \times 10^{-3}\, \,T_{\textrm{SM}}$.
The short-dashed black line near the red star 
represents the nearly-virial trajectory of the smallest fragments after the point of last fragmentation,
due to surface cooling.
\textbf{Dashed and dotted gray:} Same as above,
but for different choices for initial conditions 
$T_{e_D} = 5 \times 10^{-4}\, \,T_{\textrm{SM}}$ (dotted gray), 
and
$T_{e_D} = 0.5\, \,T_{\textrm{SM}}$ (dashed gray).
In the former case, 
the temperature-density trajectory merges with the solid black contour at high densities,
so the dependence on the initial conditions is washed out in the later stages of halo evolution.
\textbf{Solid thin gray:} Contours of Jeans mass in units of solar masses.
The contours have a slope $d\log n_{e_D}/d\log T_{e_D}=3$ at low densities where kinetic pressure dominates, 
but become temperature independent at high densities where the dark-photon repulsive force is the main source of pressure.
\textbf{Green:} Contour of dark-electron free-fall time being equal to the Bremsstrahlung cooling time 
(``cooling equals heating").
\textbf{Blue:} Region of cooling time being longer than the Universe's age (inefficient cooling).
\textbf{Red:} Region where Jeans-sized fragments are optically thick.
We set the optically thick boundary at the point of inverse Bremsstrahlung reabsorption,
$\ell_{\gamma_D}^{\textrm{abs}}=\sqrt{N_{\textrm{sc}}} \, \lambda_J/2$,
where $N_{\textrm{sc}}$ accounts for Compton rescatterings.
}
\label{fig:three}
\end{center}
\end{figure}
%%%%%%%%%%
% END FIGURE
%%%%%%%%%%

We can now determine the trajectory of the temperature and number density $n_{e_D}\equiv\rho_{e_D}/m_{e_D}$ of the dark-electron clumps 
as they collapse by using expressions Eqs.~\eqref{eq:contour2}, \eqref{eq:tdefinition},  \eqref{eq:tdefinition2}, and~\eqref{eq:coolingrate}, fixing the initial condition for the dark-electron gas density at turnaround by Eq.~\eqref{eq:ics}, and choosing a dark-electron halo temperature. 
The temperature-density trajectory for a choice of model parameters $m_{e_D}=1\,\textrm{GeV}, m_{\gamma_D}=100 \, \textrm{eV}, \alpha_D=1/10$ is presented in Fig.~\ref{fig:three} (black solid line).
In the figure, 
we concentrate on a dark-electron clump embedded in the Milky-Way, 
which we take to have a total mass $M_{\textrm{MW}}=10^{12} \, M_{\odot}$.
We also set the dark-electron to dark-matter fraction to $f=1\%$, 
so the corresponding mass of the dark-electron Milky-Way halo is $M_{\textrm{MW}}^{e_D}=10^{10} \, M_{\odot}$ (c.f. Eq.~\eqref{eq:halomass}).
On the other hand, and as we will see shortly,
the precise initial conditions for the temperature and even for the density are not important for our results.
For concreteness, for the collapse trajectory in the black solid line in Fig.~\ref{fig:three} we just take the dark-electron halo temperature to be a typical ``cosmological temperature" $T_{e_D} = 5 \times 10^{-3}\, \,T_{\textrm{SM}}$. 

From the figure, we clearly see that the collapse of the dark-electron clumps proceeds in three stages.
Chronologically,
the three stages are:
%%%%%%%%%
\begin{enumerate}
\item \textbf{Adiabatic free-fall}: 
After Hubble decoupling (at a density given by Eq.~\eqref{eq:ics}), 
the dark-electron overdensities free-fall into the gravitational potential set mostly by the main cold-dark matter component at the free-fall time Eq.~\eqref{eq:freefalltime}. 
CDM particles pass through the center of the halo without interacting and
eventually settle into an NFW-like halo.
Dark electrons, 
on the other hand,
are collisional particles on galactic scales (if the dark-photon mass is below the threshold value Eq.~\eqref{eq:photonmassconstraint})
so after a period of free fall they accumulate towards the center of the perturbation, since they transform part of their gravitational energy into thermal energy.
Near the center of the perturbation where they become dense, they also  become the dominant source of gravitational potential (where other dense baryonic clumps may be neglected).
For our analysis it is not particularly important which one is the dominant source of gravitational potential in this initial stage of collapse.
The reason is that in this stage, and as long as the halo free-falls, in Eq.~\eqref{eq:contour2} $t_{\textrm{collapse}} \ll t_{\textrm{cooling}}$ 
so the slope of the collapse trajectory is just $d\log T_{e_D}/ d\log n_{e_D} \simeq 2/3$.\footnote{This corresponds simply to the adiabatic condition for a monoatomic ideal gas.}
Now, 
the regions where the cooling time is longer than the age of the Universe (inefficient cooling) are shown in blue.
Initially, we see that collapse proceeds adiabatically,
since the halo temperature is lower than the dark-photon mass,
so Brems\-strahlung cooling is strongly suppressed.
Cooling starts becoming efficient when the dark electrons become sufficiently hot and dense.
On the other hand, as the gas collapses the kinetic pressure increases,
so we see that the temperature-density trajectory moves into regions of higher Jeans masses,
shown in Fig.~\ref{fig:three} in solid gray contours in units of solar masses.
It is likely that this stage is then accompanied by mergers of clumps into a unique, larger and roughly homogeneous galactic halo.

\item \textbf{Nearly virialized contraction}: 
The increase in temperature in the dark-electron halo leads eventually to halo virialization by kinetic pressure.
This represents a transition from the free-fall regime into the nearly-virialized-contraction regime, 
c.f.~Eq.~\eqref{eq:tdefinition2}.
To match these two transitions, 
we simply intersect the free-fall trejectory with the virial line set by $M_{\textrm{MW}}^{e_D}=m_{J}$.
At this stage,
a roughly homogeneous dark-electron galactic halo continues to collapse in a quasi-static configuration close to the virialized state, 
at a timescale set by the cooling rate.
The halo will efficiently collapse if the cooling time is shorter than the age of the Universe.
As pointed out above, 
Brems\-strahlung cooling is efficient at high densities and temperatures,
so for collapse to continue at this stage, the halo must be hot and compact at virialization.
In particular, 
the dark-electron gas must achieve a temperature larger than the dark-photon mass to trigger Bremsstrahlung.
To assess if high-enough temperatures and densities are achieved to trigger Bremsstrahlung, 
we must first investigate the process of halo virialization in more detail.

\begin{itemize}[leftmargin=*]
\item \textit{Details of halo virialization:}
While up to now our simplified analysis captures the overall features of halo collapse, 
including a period of free fall followed by quasi-static virialized contraction due to cooling, 
it does not capture an important element, 
which is the presence of shocks~\cite{1977MNRAS.179..541R}. 
In a realistic analysis, collisional particles have indeed a period of adiabatic contraction, 
but that ends abruptly before the gas contracts significantly, 
by the creation of a shock that expands from the center of the perturbation. 
Inside the shock the gas settles into a virialized state (approximately isothermal and with radial density profile $\sim r^{-2}$~\cite{1985ApJS...58...39B}), 
while outside the shock the gas free falls into the shock boundary,
so the adiabatic free fall and nearly-virialized-contraction stages of collapse actually coexist. 
Importantly, 
in the process of the shock the dependence of the initial ``cosmological'' temperature of the dark electrons is lost, 
and dark electrons inside the shock are simply heated up to their virial temperature~\cite{1985ApJS...58...39B}.
Despite its simplicity, 
our elementary analysis is in fact able to partially capture the washout of the initial ``cosmological" temperature at virialization. 
This is  illustrated by the dotted-gray trejectory that shows an alternative (colder) choice of the dark-electron halo initial temperature,
$T_{e_D} = 5 \times 10^{-4}\, \,T_{\textrm{SM}}$.
While the corresponding collapse trajectory differs at the initial stage of collapse with our original trajectory (in solid black),
both trajectories later merge into a unique asymptotic collapse trajectory after virialization.
On the other hand, 
a limitation of our analysis can be discovered by considering instead a ``hot" initial condition $T_{e_D} = 0.5 \, \,T_{\textrm{SM}}$ (dashed gray).
In this case,
it would seem that the dark-electron halo virializes quickly at small densities, where cooling is inefficient,
so the halo ends up in the hypothetical final state shown by the black circle.
This is unphysical: as stated above the final result in a realistic analysis may not depend on the initial ``cosmological" temperature.
In a more realistic analysis accounting for shock heating,
and as long as the dark-electron temperature reaches the dark-photon mass,
cooling will proceed (at least near the center of the perturbation that is dense, where Bremsstrahlung is most efficient).
In this case, Bremsstrahlung cooling is triggered and the stage of nearly virialized contraction is ensured to happen.
If, on the other hand, 
the temperature of the dark-electron halo does not reach the dark-photon mass after shock heating, 
the dark-electron halo reaches its final state as a virialized and roughly isothermal sphere with the typical $\sim 1/r^2$ profile~\cite{1985ApJS...58...39B}.\footnote{This picture may be affected by the gravothermal evolution of the halo~\cite{1968MNRAS.138..495L}. The authors in 
\cite{Pollack:2014rja,Essig:2018pzq} find that under certain assumptions for the initial distribution, a subdominant component of dark matter with elastic self interactions undergoes gravothermal collapse. Studying this in detail is beyond the scope of this paper.}
The important question is then if the halo ever reached temperatures above the dark-photon mass.
Assuming spherical symmetry and in the absence of angular momentum,
the after-shock temperature has been calculated in~\cite{1985ApJS...58...39B} (or may be read off more directly from~\cite{Wang:2008zzi,Nelson:2015cda,Barkana:2000fd,Bromm:2013iya}), and it is given by 
\begin{equation}
T_{\textrm{shock}}^{e_D} \simeq 3 \times 10^{-3} 
(1+z_{\textrm{ta}}) \, \bigg[\frac{m_{e_D}}{1\, \textrm{MeV}}\bigg]\bigg[\frac{M_{\textrm{halo}}^{e_D}}{10^{10} \, M_{\odot}}\bigg]^{2/3}\, \textrm{eV}
\quad ,
\label{eq:shock}
\end{equation}
where $z_{\textrm{ta}}$ is the halo-turnaround redshift, which for the Milky Way is $z_{\textrm{ta}} \simeq 1.5$ (see Section~\ref{sec:transition}). 
This temperature has only a mild radial dependence $\sim r^{-1/4}$~\cite{1985ApJS...58...39B}.
In our final results in Section~\ref{sec:smallest}, 
we will indicate the regions of parameter space in which $T_{\textrm{shock}}^{e_D}  \geq m_{\gamma_D}$,
\textit{i.e.}, when the shock  temperature is high enough to trigger Bremsstrahlung.
In this case collapse via cooling in the nearly virialized stage is ensured.
Note however, 
that it is possible that the after-shock temperature may still increase due to other mechanisms, 
such as halo mergers~\cite{1992ApJ...389....5T},
or gravothermal collapse of the inner core~\cite{Pollack:2014rja,Essig:2018pzq},
leading to further collapse even if shocks cannot trigger Bremsstrahlung.
In any case, 
as long as Bremstrahlung was triggered,
we may proceed to the last stage of collapse, which is the period of fragmentation.
\end{itemize}
\item \textbf{Fragmentation}: 
Quasi-static collapse near the virialized state leads to an increase in dark-electron density,
and it also leads to an increase in the Brems\-strahlung cooling rate.
At large enough densities, 
the cooling time becomes comparable to the free-fall time.
In Fig.~\ref{fig:three}, we show in green the contour of cooling time being equal to the free-fall time, $t_{\textrm{cooling}} = t_{\textrm{ff}}$,
which represents a trajectory of ``cooling equals heating''.
At this stage,
the gas collapses at the free-fall time and the collapse contour follows closely the ``cooling equals heating" line.
This result is quite intuitive: 
since cooling is extremely efficient in this latter stage of collapse, 
as soon as the halo collapses and converts gravitational energy to thermal energy,
this energy is released via Bremsstrahlung.
Importantly,
the halo ends up following the ``cooling equals heating" line \textit{regardless of the details of all the previous stages of collapse},
so as long as fragmentation started, we do not have to worry anymore about shocks, mergers, or any other complicated processes. 
The ``cooling equals heating'' trajectory leads to a decrease in the Jeans mass,
so the dark-electron halo,
which was roughly homogeneous at virialization,
starts dividing into smaller clumps or \textit{fragments} \footnote{In astronomy literature the condition $t_{\textrm{cooling}}\leq t_{\textrm{ff}}$ is called the Rees-Ostriker-Silk criterion for halo fragmentation~\cite{Bromm:2013iya}.}.
The trajectory follows the ``cooling equals heating" contour up to the point of last fragmentation, 
shown with a red star.
\end{enumerate}

Fragmentation stops for one of two possible reasons.
The first possibility,
is that as the density becomes too large and the temperature decreases too much,
cooling becomes inefficient. 
The reason is that at high densities the dark-electron gas becomes either optically thick 
(as happened to be the case in Fig.~\ref{fig:three}) or the temperature falls below the dark-photon mass so that 
Brems\-strahlung is exponentially suppressed.
For the selected model parameters in Fig.~\ref{fig:three},
the point of last fragmentation (red star) corresponds to a fragment with a mass of order $\sim 50 \, M_{\odot}$.
This \textit{minimal fragment} is the analogue of a conventional protostar in the baryonic sector. 
Note that these minimal fragments are not stable, 
since even if they are optically thick,
they may further collapse by cooling from the surface.
We come back to this issue in the next section.

The second possibility for the end of fragmentation is shown in Fig.~\ref{fig:four},
where we choose a smaller dark-photon mass and larger dark-electron mass with respect to the choice in Fig.~\ref{fig:three},
$m_{e_D}=10 \, \textrm{GeV}$,  $m_{\gamma_D}=0.1 \, \textrm{eV}$.
In this case, at high densities and due to the smaller dark-photon mass,
the pressure in Eq.~\eqref{eq:pressure} is dominated by the dark-photon repulsive force.
As a result, 
the Jeans mass becomes independent of the dark-electron temperature and simply increases monotonically with density,
so cooling does not lead to a decrease of the Jeans mass and further fragmentation.
In this scenario, 
the minimal fragments reach their final state right at the end of fragmentation, 
being stabilized by the dark-photon repulsive force.
For the choice of model parameters leading to the collapse trajectory in Fig.~\ref{fig:four}, 
the minimal fragments have a typical mass of $\sim 10^6$ solar masses. 

%%%%%%%%%%
% BEGIN FIGURE
%%%%%%%%%%
\begin{figure} [t!]
\begin{center}
\includegraphics[width=10cm]{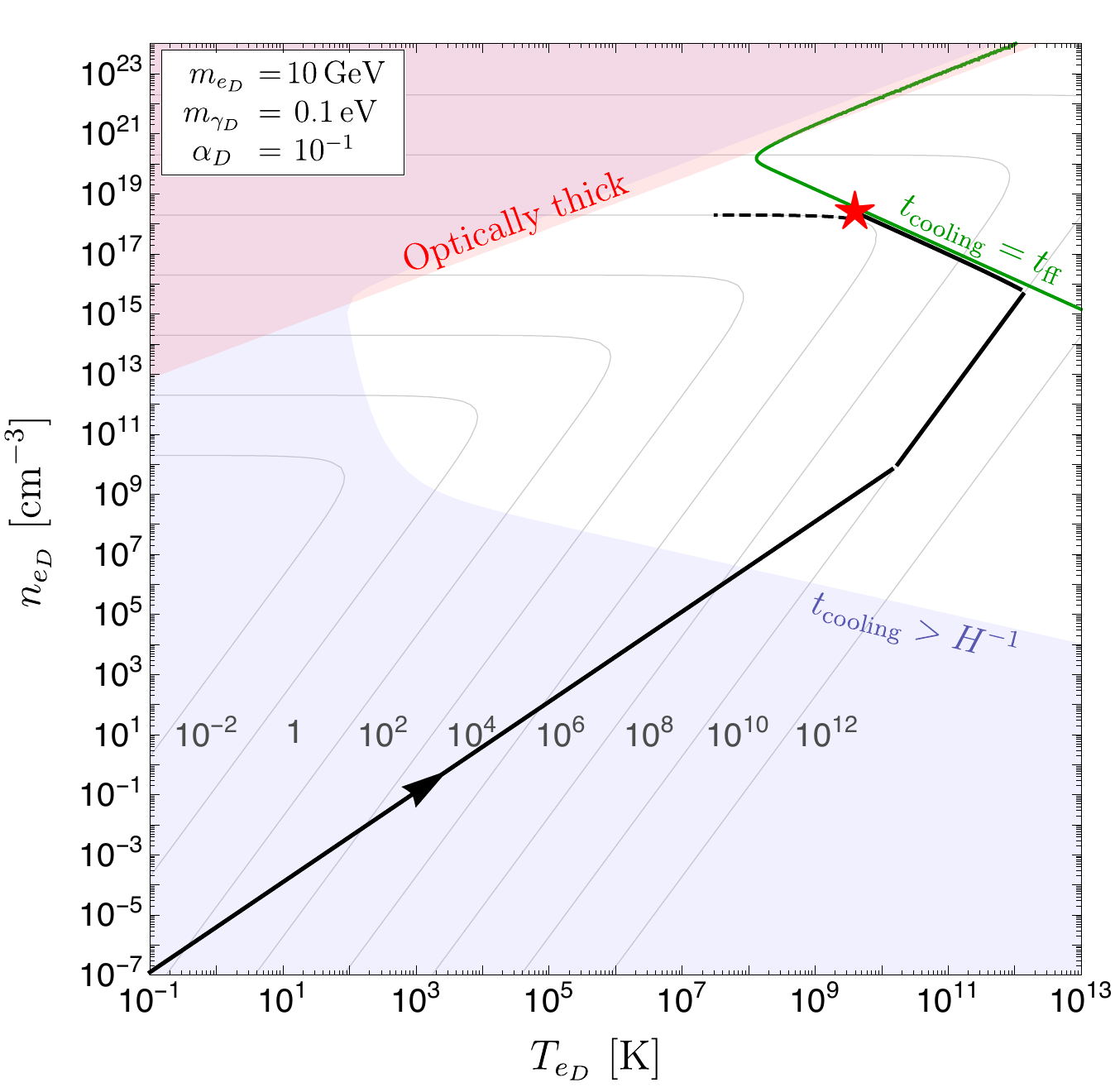}
\caption{Same as Fig.~\ref{fig:three} but with a smaller choice of dark-photon mass, $m_{\gamma_D}=0.1 \, \textrm{eV}$ and larger dark-electron mass, $m_{e_D}=10 \, \textrm{GeV}$.}
\label{fig:four}
\end{center}
\end{figure}
%%%%%%%%%%
% END FIGURE
%%%%%%%%%%

We acknowledge that our study of the non-linear evolution of the dark-electron halos has important limitations.
For instance, 
it does not describe in detail the process leading to halo virialization, mergers and halo shape.
All these elements cannot be analyzed in detail without a numerical simulation.
However,
we stress that the typical size of the minimal fragment is rather independent of the fine details of the non-linear dark-electron halo evolution (which could only be captured by a full numerical simulation).
The reason is that the size of the minimal fragment is set by the temperature-density trajectories,
and these trajectories have the same asymptotic form:
they converge to the ``cooling equals heating" trajectory.
The ``cooling equals heating" trajectory depends uniquely on the particle model parameters, 
\textit{i.e.}, on the dark-electron and dark-photon masses and the dark fine-structure constant.
So as long as the conditions leading to the beginning of fragmentation studied throughout this work are satisfied,
dark-electron halos with different assembly and collapse histories lead to the same minimal fragments.
This allows us to give robust predictions for the typical size of the astronomical objects formed by fragmentation in the dark-electron sector.
Note that this also means that the typical size of the dark-electron smallest fragments is \textit{universal},
\textit{i.e.} it should be roughly the same in other galaxies that are more or less massive than the Milky Way.
\footnote{This is also true for the baryonic-sector fragments. The typical mass of baryonic stars is roughly independent of the mass of the host galaxy.} 
We dedicate the next section to providing a detailed study of the size of the smallest dark-electron fragments.

\subsection{Mass and compactness of the exotic compact objects}
\label{sec:smallest}
We now obtain the mass and compactness of the minimal fragments
as a function of the dark-electron model parameters, 
by obtaining the temperature-density trajectories for the dark-electron gas and the last point of fragmentation (as in Figs.\ref{fig:three} and \ref{fig:four})
for different dark-electron and dark-photon masses and different choices of the dark-sector fine-structure constant.
The results are presented in Fig.~\ref{fig:five} for fine-structure constant $\alpha_D=10^{-1}$ (top), $\alpha_D=10^{-2}$ (middle) and $\alpha_D=10^{-3}$ (bottom). 
In the figure, black contours indicate the mass of the fragments in solar masses and blue contours, 
their compactness. 
Note that we concentrate on a range of dark-electron masses $m_{e_D} \geq 1 \, \textrm{MeV}$ consistent with growth of perturbations during the linear regime,
as discussed in Section~\ref{sec:lineardynamics}.
We indicate in shaded red
regions where dark electrons are collisionless at turnaround (c.f.~Eq.~\eqref{eq:photonmassconstraint}).
In these regions we do not expect the dark electron to become hot, compact, or fragment,
and we expect instead that it resembles a cold dark matter halo.
In addition, 
we show with a red-dashed line the \textit{maximal} dark-photon mass for which Bremsstrahlung cooling can be triggered by shock heating.
We obtain this mass from the condition $T_{\textrm{shock}}^{e_D} \geq m_{\gamma_D} $ and using Eq.~\eqref{eq:shock} with $M_{\textrm{halo}}^{e_D}=10^{11}\, M_{\odot}$,
which is typically the maximal dark-electron halo mass consistent with $f\leq 10\%$ (c.f. Eq.~\eqref{eq:halomass}). 
Since Bremsstrahlung cooling is needed for fragmentation to start, 
in the absence of additional halo heating it is likely that the halo efficiently fragmented only in regions of parameter space \textit{below} the red-dashed line.
However,
to account for the possibility that the halo may have been heated above its shock temperature by some other dynamical process such as gravothermal collapse,
we also present the size of the minimal fragments in the regions of parameter space above the red-dashed line.
Bearing this important observation in mind,
we now proceed to discuss the properties of the minimal fragments across all the parameter space presented in Fig.~\ref{fig:five}.

We start by discussing the minimal fragment's mass.
First, we point out that the masses of the minimal fragments in Fig.~\ref{fig:five} are a \textit{lower limit} on the mass of exotic compact objects in the dark sector, 
since such compact objects may still grow due to accretion and mergers after the point of last fragmentation.
However, 
in the baryonic sector the mass of the minimal fragments gives a good order of magnitude estimate of the typical mass of stars~\cite{doi:10.1093/mnras/176.2.367},
so we also expect the results in Fig.~\ref{fig:five} to be a faithful representation of the mass of exotic compact objects in our dark sector.
By comparing the top, middle, and bottom panels in Fig.~\ref{fig:five}, we see that generically fragmentation is most efficient (\textit{less massive} fragments are formed) 
for larger values of the fine-structure constant, 
since this leads to efficient cooling. 
We also see that fragmentation is efficient when \textit{both} the dark-photon and dark-electron masses are large, 
\textit{i.e.}, 
in the upper-right quadrant of all the plots in Fig.~\ref{fig:five}.
The reason is that large dark-photon masses lead to a suppression of the dark-photon repulsive force and a reduction of the Jeans mass. 
On the other hand, 
large dark-electron masses lead to an increase of the dark-photon absorption mean free path Eq.~\eqref{eq:libs}, 
so the material remains optically thin favoring cooling and fragmentation.
In particular, 
for $m_{\gamma_D} \gtrsim 1\, \textrm{keV}$ and $m_{e_D} \gtrsim 1\, \textrm{GeV}$, 
the halo fragments into solar-mass sized dark ``asymmetric stars.'' 

%%%%%%%%%%
% BEGIN FIGURE
%%%%%%%%%%
\begin{figure} [!]
\begin{center}
\begin{minipage}[c]{0.55\textwidth}
\includegraphics[width=8cm]{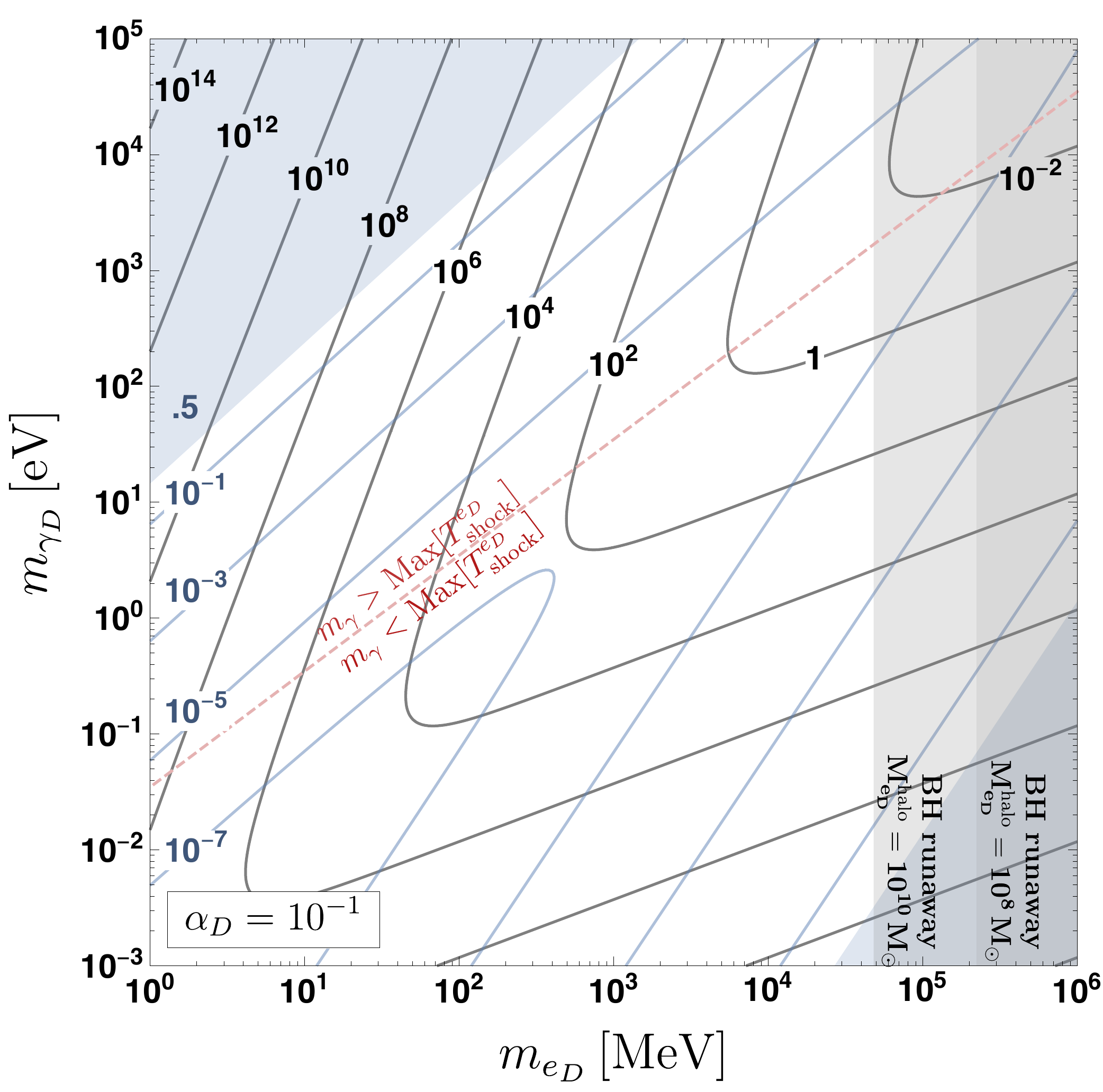}
\includegraphics[width=8cm]{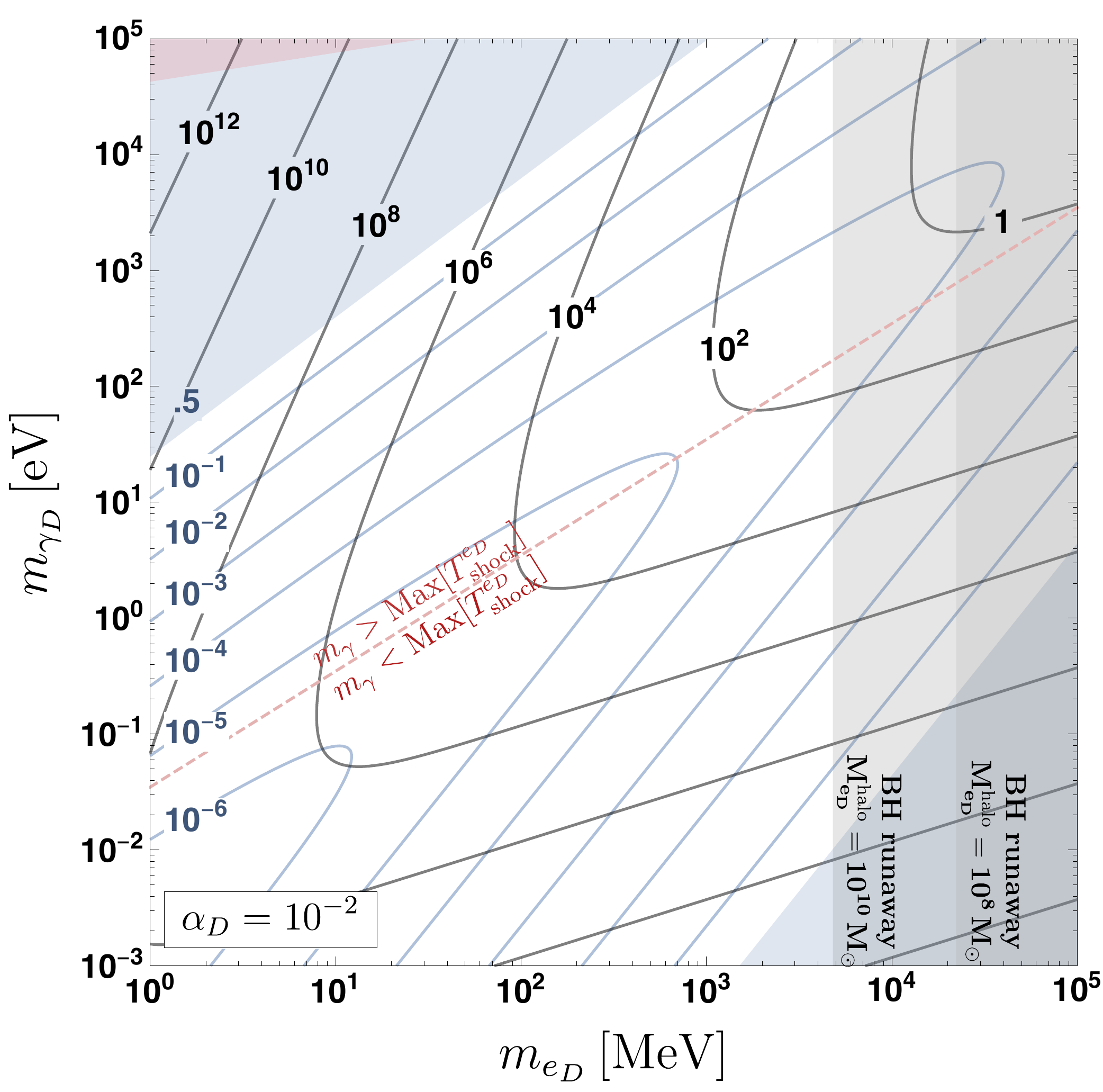}
\includegraphics[width=8cm]{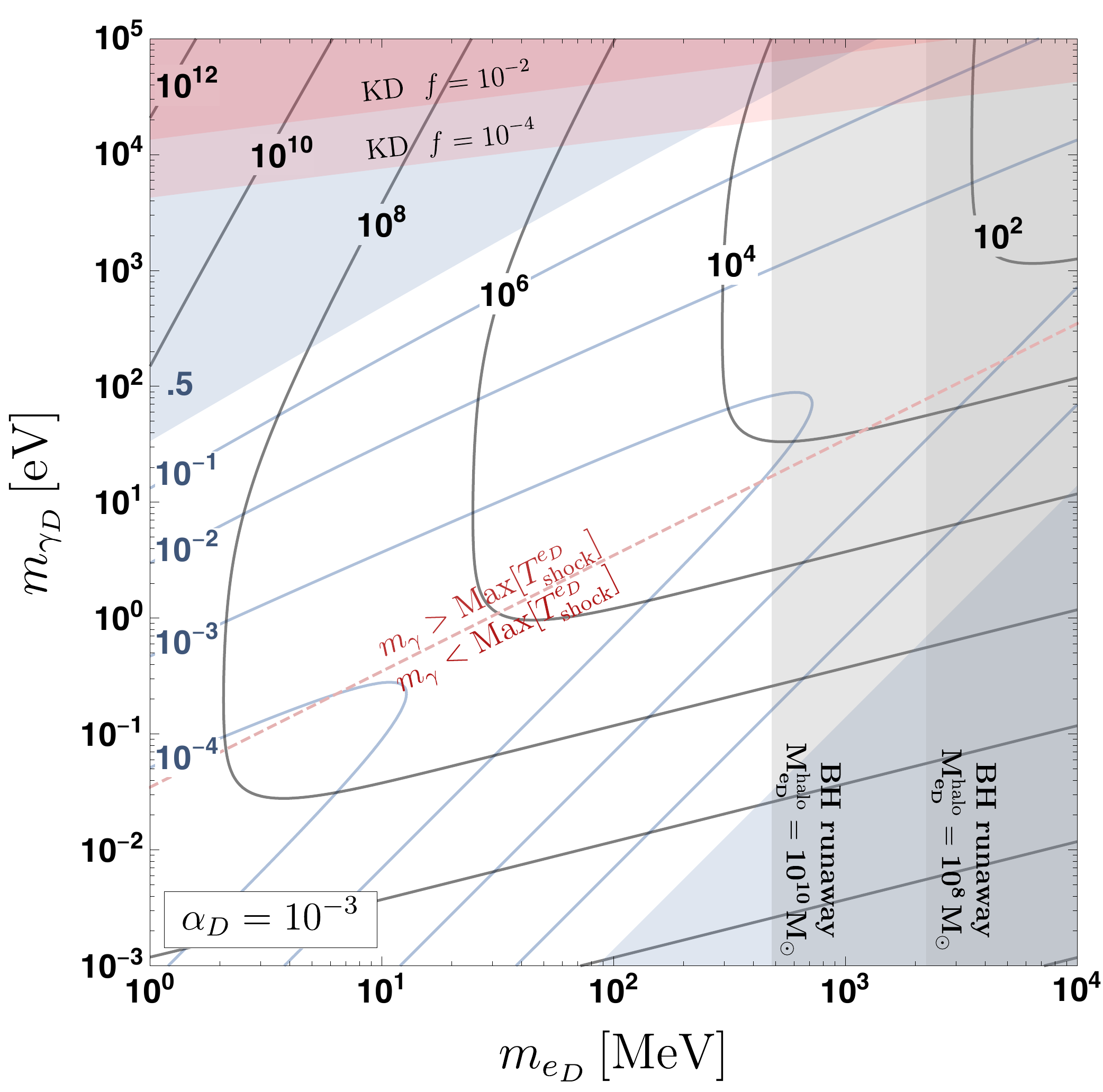}
  \end{minipage}\hfill
  \begin{minipage}[c]{0.45\textwidth}
    \caption{
\textbf{Black contours:} Mass of the dark-electron exotic compact objects formed via fragmentation of a dark-electron halo, in units of solar mass, 
as a function of the dark-electron and dark-photon masses.
The dark fine-structure constant has been set to $\alpha_D=10^{-1}$ (top), $\alpha_D=10^{-2}$ (middle) and $\alpha_D=10^{-3}$ (bottom).
\textbf{Blue contours:} Compactness of the above objects.
\textbf{Shaded blue:}
Regions where the dark-electron exotic compact objects are black holes (compactness $C_{\textrm{BH}}=1/2$).
\textbf{Shaded gray and red:} 
Regions where no fragmentation occurs. 
In gray, 
no fragmentation occurs since the whole dark-electron halo runs away into a black hole before it can fragment.
We plot these gray regions for two choices of dark-electron halo mass.
In red, 
we show the regions where dark electrons are collisionless (\textit{i.e.}, kinetically decoupled (KD)) during linear growth of perturbations (c.f.~Eq.~\eqref{eq:photonmassconstraint}),
so instead of forming a compact dark-electron halo that fragments, 
dark electrons settle in an NFW-like halo typical of CDM. 
We show these regions for two choices of the dark-electron to dark-matter ratio. 
In dashed-red, we show the \textit{maximal} dark-photon mass for which Bremsstrahlung cooling can be triggered by shocks within the dark-electron halo, assuming $f=10\%$ and $M_{\textrm{halo}}=10^{12}\, M_{\odot}$ (see text for details). 
    } 
    \label{fig:five}
     \end{minipage}
\end{center}
\end{figure}
%%%%%%%%%%
% END FIGURE
%%%%%%%%%%

With the dark-electron mass held fixed,
a too-large dark-photon mass generically leads to the formation of more-massive compact objects,
since it leads to an exponential suppression of the Bremsstrahlung cooling rate. 
For a fixed dark-photon mass, 
a too-large dark-electron mass also disfavors fragmentation, since it suppresses the cooling rate,
so it postpones fragmentation to large densities where the Jeans mass can be enhanced by the dark-photon repulsive force.
In fact, 
if the dark-electron mass is too large, 
efficient fragmentation is postponed to extremely large dark-halo densities.
At some point, 
the densities required for fragmentation are so large that the whole dark-electron halo runs away into a black hole before fragmentation starts.
In Fig.~\ref{fig:five},
we indicate the halo runaway regions in gray,
for two different choices of dark-electron halo mass.

Our results are not reliable close to the black-hole runaway regions,
since it is possible that once angular momentum is included,
the dark-electron halo will be stabilized before reaching the high densities required for fragmentation.
Nevertheless, 
an interesting situation arises if the dark-electron halo has no angular momentum and indeed runs away into a black hole without fragmenting. 
In this case, 
the mass of such black hole is fixed by the dark-electron to dark-matter ratio $f$ to be the corresponding fraction of the host galaxy's mass, $M_{\textrm{BH}}=f M_{\textrm{galaxy}}$.
If $f\sim 10^{-6}$, 
this scenario could account for the super-massive black hole in the center of the Milky Way (although other known mechanisms exist to 
explain its formation~\cite{vdb}). 
Moreover, a prediction of this scenario is that more massive galaxies would contain more massive black holes at their center, 
the relation between their masses being close to linear,
consistent with current observations~\cite{Bandara:2009sd}.
In addition,
this scenario could explain the existence of supermassive black holes at high redshift~\cite{dedcb98aa4b4415390d0dde79a27bce4,0004-637X-790-2-145,2011Natur.474..616M}. 
We postpone a more detailed discussion to the problem of angular momentum to Section~\ref{sec:angularmomentum}.

Fragmentation tends to be inefficient for small dark-electron and dark-photon masses, \textit{i.e.}, in the lower-left quadrant of all the plots in Fig.~\ref{fig:five}. 
In these regions of parameter space, 
the small dark-photon mass leads to a large enhancement of the dark-photon repulsive force,
while the small dark-electron mass overly enhances the dark-photon inverse bremsstrahlung reabsorption rate,
so the halo does not efficiently cool. 
For dark-photon masses below roughly $10^{-3} \, \textrm{eV}$ and dark-electron masses below roughly $1\, \textrm{MeV}$ the halo does not fragment,
and the end-result is a large and virialized dark-electron halo devoid of substructure and without exotic compact objects.  

The second important property of an exotic compact object is its compactness,
which is an indicator of its density,
and is defined as
\begin{equation}
C
\equiv
\frac{MG}{R}
\quad ,
\label{eq:compactnessdef}
\end{equation}
where $M$ is the mass of the fragment and $R$ is its radius.
$C_{\textrm{BH}}=1/2$ for a non-rotating black hole, 
$0.13 \leq C_{\textrm{NS}} \leq 0.23$ for a neutron star~\cite{Lattimer:2006xb}, and
$C_{\odot} \simeq 10^{-6}$ for the Sun.
The density of the minimal fragments may be obtained straightforwardly by finding the last point of fragmentation as in Fig.~\ref{fig:three} and~\ref{fig:four}. 
However, as discussed in the previous section, this is not the density of the final exotic compact object. 
Once fragmentation stops, 
the smallest fragments may still collapse by cooling from the surface, 
until they become supported by the dark-photon repulsive force, by dark-electron degeneracy, or until they runaway into a black hole, 
at which point they achieve their maximal compactness (at a temperature $T_{e_D} \simeq m_{\gamma_D}$, since below this temperature Bremsstrahlung shuts off).
Surface cooling is efficient if the corresponding cooling time is shorter than the age of the Universe.
To estimate if fragments had enough time to cool into their final equilibrium state,
we compute the surface cooling time using the black-body cooling rate,
and compare it with the age of the Universe.
We find that in all the parameter space in Fig.~\ref{fig:five}, fragments had enough time to reach their final equilibrium state, 
\textit{i.e.}, their maximal compactness.
Finally, to estimate the fragment's maximal compactness we neglect degeneracy pressure, 
and calculate the object's radius when stabilized by the dark-photon repulsive force.
We leave the details of the derivation to Appendix~\ref{app:three}.
The dark-electron fragments maximal compactness is
%%%
\begin{equation}
C
=
\frac{ m_{e_D} m_{\gamma_D} M}
{\pi}
\bigg(
\frac{G^3}
{ \alpha_D}
\bigg)^{1/2}
\quad ,
\label{eq:compactness}
\end{equation}
%%%
where $M$ is the mass of the fragment.
While expression Eq.~\eqref{eq:compactness} is only an estimate of the object's exact compactness, 
we have checked that it gives a good approximation by comparing with the exact results in~\cite{Kouvaris:2015rea,Gresham:2018rqo},
except for values close to the black-hole compactness, $C_{\textrm{BH}}=1/2$. 

From Fig.~\ref{fig:five}, 
we conclude that collapse and fragmentation of the dark-electron halo leads to highly compact objects, typically more compact than the Sun. 
These objects range in mass from a few to millions of solar masses.
As an example of the former situation,
consider the case $\alpha_D=10^{-1}$, $m_{\gamma_D}=0.04 \, \textrm{eV}$, $m_{e_D}= 1 \,\textrm{GeV}$.
In this scenario,
the typical mass of the smallest fragment is $\sim 10^{6} \, M_{\odot}$,
and its maximal compactness is roughly an order of magnitude larger than the Sun's compactness, $C\simeq 10^{-5}$.
As an example of the latter scenario,
take $\alpha_D=10^{-1}$, $m_{\gamma_D}=100 \, \textrm{eV}$, $m_{e_D}= 10 \,\textrm{GeV}$ . 
In this case, 
the fragments have a mass of $\sim 1 \, M_{\odot}$ and a compactness close to that of the Sun, $C\simeq 10^{-6}$ 
In addition, 
in the whole blue shaded region the fragments are expected to be black holes.

Finally, 
the remaining crucial property of the exotic compact objects is their abundance. 
Differently from the mass and compactness of the fragments,
their abundance does depend on the dark-electron halo mass,
or equivalently through Eq.~\eqref{eq:halomass} on the dark-electron to dark-matter ratio $f$.
Assuming that the whole dark-electron halo goes into the minimal fragments,
one may obtain an upper limit for the number of fragments in the Milky Way halo
given by~\cite{1978MNRAS.184...69L}
\begin{eqnarray}
\nonumber 
N_{\textrm{max}}
&=&
\frac{M_{\textrm{MW}}^{e_D}}{M}
\\
&=& 
f \, \frac{M_{\textrm{MW}}}{M} 
\quad ,
\end{eqnarray}
where $M$ is the mass of the fragment and in the second equality we made use of  Eq.~\eqref{eq:halomass} to relate the dark-electron halo mass with the Milky Way mass $M_{\textrm{MW}}=10^{12} \, M_{\odot}$ and $f$.
For instance, taking $f=1\, \%$ and $\alpha_D=10^{-1}$, $m_{\gamma_D}=100 \, \textrm{eV}$, $m_{e_D}=10 \, \textrm{GeV}$, 
we find that the Milky Way would contain roughly $10^{10}$  asymmetric dark matter stars with a mass $M \sim \, M_{\odot}$ that are roughly 
as compact as the sun $C\simeq 10^{-6}$. 
This is roughly one solar-mass sized asymmetric dark-matter star per ten baryonic stars.

\subsection{A dark-electron halo with angular momentum}
\label{sec:angularmomentum}
In the previous discussion we have neglected the angular momentum of the dark-electron halo.
The angular momentum of the galaxy may arise from tidal torques during linear evolution~\cite{1969ApJ...155..393P,1970Ap......6..320D} 
or from mergers with other halos~\cite{2002MNRAS.329..423M}. 
In all cases, 
since dark-matter, baryons, and dark electrons in the very early stages of formation of a galactic halo behave in essentially the same way, 
it is likely that the specific angular momentum of all components is similar~\cite{1983FCPh....9....1E}. 

If the dark-electron halo rotates, 
collapse in directions perpendicular to the rotation axis is resisted by centrifugal forces,
but dark electrons may still collapse towards the rotation plane. 
During the initial stage of the dark-electron halo collapse,
a cold dark-electron halo free-falls and may not loose energy via Brems\-strahlung,
since it is too cold to emit the massive dark photons.
As a consequence,
the problem of dark-electron halo collapse in the presence of angular momentum 
is not the same as the problem of baryonic disk formation, 
which is accompanied by energy loss due to cooling.
Instead, 
the collapse of the rotating dark-electron halo closely resembles the problem of adiabatic collapse of an homogeneous rotating gas sphere. 
This problem has been studied both analytically~\cite{1962PCPS...58..709L} and numerically~\cite{1977PThPh..58..802K,1996A&AS..115..573A}.
The final result is that the rotating gas sphere collapses into a disk-like shape.
Also, a disk may be formed by accretion
of dark electrons into the baryonic potential~\cite{Read:2008fh,Purcell:2009yp,2009MNRAS.397...44R}.
In the presence of mergers, though, the disk may be disrupted~\cite{Ghalsasi:2017jna,1992ApJ...389....5T}, and the final halo distribution may be spheroidal.

In any case,
whether the final shape of the dark-electron halo is a rotating sphere or a rotating disk,
as discussed in the previous sections, the properties of the minimal fragments depend only on microscopic parameters,
as long as fragmentation happened within the halo. 
As a consequence, the inclusion of halo angular momentum may change the overall distribution of the dark-electron gas in the galaxy, 
but does not change the typical size of the exotic compact objects. 

\FloatBarrier 

\section{Observational signatures}
\label{sec:five}
In this section, we briefly comment on the experimental opportunities to discover the dark-electron sector.
A detailed analysis of the dark-electron signatures will be presented in future work.
 
Even if dark electrons are a subdominant component of the dark matter and do not necessarily interact with the Standard Model,
dark-electron exotic compact objects generically lead to interesting gravitational signals,
similar to the ones of massive compact halo objects (MACHOs).
Due to the high compactness of the dark-electron fragments (see Fig.~\ref{fig:five}) for all dynamical and lensing constraints,
they are point-like objects.
For fragment masses $\lesssim 10 \, M_{\odot}$, 
the amount of dark-electron fragments between Earth and the Large Magellanic Clouds can be efficiently constrained
by microlensing~\cite{Tisserand:2006zx,Griest:2013aaa} to be roughly at the percent level of the total dark-matter content.
Fragment masses $10 \, M_{\odot} \lesssim M\lesssim 10^5 \, M_{\odot}$,
may be constrained by the disruption of the Eridanus II star cluster~\cite{Brandt:2016aco}.
Fragment masses $10^5 \, M_{\odot} \lesssim M\lesssim 10^{10} \, M_{\odot}$ are strongly constrained from the amount of dark mass in the galactic center,
where they fall in by friction~\cite{Carr:1997cn}.
In this range, constraints on the dark-electron fraction are as strong as one part in $10^4$.
All these constraints depend on the distribution of dark-electron fragments within the galaxy,
and they may easily vary by a factor of a few under different assumptions for the halo shape (for an example see~\cite{Alcock:2000ph}).
Other interesting proposals exist to detect compact objects using astrometric weak lensing with Gaia data~\cite{Prusti:2016bjo,1998astro.ph..5360D,2002MNRAS.331..649B,VanTilburg:2018ykj}. Also, a dark-disk may be constrained using stellar kinematics~\cite{Schutz:2017tfp,Buch:2018qdr}.
In addition, 
since it is likely that dark-electron gas accretes into deep baryonic gravitational potentials,
accumulation of dark electrons in the Sun may be constrained by helioseismology~\cite{Fan:2013bea,2010PhRvD..82j3503C}.
Finally, 
high-resolution probes of the CMB may also provide insight into dark matter on small scales~\cite{Nguyen:2017zqu}.

Another interesting direction is to look for exotic compact object binaries via their gravitational-wave emission signatures~\cite{Giudice:2016zpa,Cardoso:2016oxy,Maselli:2017vfi,Urbano:2018nrs,Johnson-McDaniel:2018uvs}.
Note that in our analysis, the typical separation of dark-electron fragment binaries may be simply estimated from the Jeans length at the point of last fragmentation.
In our model we find  some regions of parameter space with fragments in the $\sim 1-10^2 \, M_{\odot}$ mass range as compact as a black hole, 
which could be detected at LIGO.
There are also more significant regions of parameter space where more massive black holes and solar-mass sized  asymmetric dark matter stars as compact as white-dwarfs are formed.
Such objects can be easily detected at LISA~\cite{AmaroSeoane:2012je,AmaroSeoane:2012km}. 
We point out, however,
that most of the regions of parameter space leading to solar-mass sized compact objects are above the red-dashed lines in Fig.~\ref{fig:five}.
This means that in order for such objects to be formed by fragmentation of the dark-electron halo, 
an additional heating mechanism of the halo beyond shocks is likely needed. 

An interesting interplay between the experiments above,
which rely purely on gravitational interactions,
and dark-matter direct or indirect detection experiments arises if we allow for dark-photon mixing with the Standard Model photons.
First, efficient fragmentation of the dark-electron halo gives a target for the dark-photon masses to be explored,
roughly given by $10^{-3} \, \textrm{keV}\leq m_{\gamma_D} \leq 100 \, \textrm{keV}$.
Currently, in the presence of mixing with the Standard Model photons, 
this range of dark-photon masses is strongly constrained by limits from helioscopes and star cooling~\cite{An:2013yfc,Jaeckel:2013ija}.
Due to the strong constraints on dark-photon mixing with Standard Model photons,
we do not expect our dark-electron compact objects to be bright,
but this is a direction that is still worth investigating.
Finally, since the properties of the dark-electron fragments can be straightforwardly traced back to the underlying particle-model parameters using the results in Fig.~\ref{fig:five},
an observation of a dark-sector compact object would immediately give a target for dark-matter direct- or indirect-detection experiments.

\section{Conclusions}

In this work, we present the complete history of structure formation in a dissipative dark-sector model. 
To the best of our knowledge, 
this is the first comprehensive analysis of the formation of galactic substructure and exotic compact objects in a dissipative dark-sector.
In our dissipative dark-sector, we show that small primordial density perturbations grow linearly after matter-radiation equality, 
decouple from the Hubble flow, and become dense halos. 
These halos fragment due to cooling and lead to the formation of exotic compact objects.

In the literature, the study of dark-sector compact objects has been mostly dedicated to proposing new types of compact objects,
without specifying how they formed.
Here we point out that without a full study of structure formation in the underlying dark-sector models, 
it is not possible to assess the viability of such proposals.
In our dark-sector model example, 
we find that exotic compact objects only form in specific regions of model parameter space and can only have a range of specific masses. 
We expect these conclusions to hold for other dissipative dark-sector models, 
since the regions where compact objects may be formed are selected by generic microscopic and thermodynamic properties of the model, such as the cooling rate, optical thickness, and pressure.
This has important consequences for the theory efforts aiming 
to study exotic compact objects or galactic substructure departing from the CDM paradigm.

Our work opens up several avenues for further exploration.  
First, while here we concentrated on one simplified dark-sector model, 
most of the techniques that we used can now be applied to other dissipative dark-matter models,
such as atomic dark matter.
The key observation is that once linear growth of perturbations and the beginning of halo fragmentation is ensured, 
the typical size of the halo fragments can be obtained by calculating asymptotic halo collapse trajectories. 
These trajectories are determined by microscopic dark-sector physics, 
more specifically, by the dark-sector cooling rate and the gas opacity or pressure.
Any dark matter model with a light mediator and strong interactions may lead to the formation of  interesting galactic substructure and compact objects via halo fragmentation,
but the regions of parameter space where this occurs are yet to be explored for different models.

Second, 
our dark-sector model provides strong motivation for a series of observational efforts.
Dark-sector compact objects may be discovered at high precision observatories looking for massive compact halo objects via lensing or gravitational wave detection.
A particularly interesting direction is to study the interplay between baryonic and dark-sector galactic substructure.
Since baryonic and dark matter overdensities fall into the same gravitational potentials, 
we expect that regions where baryons accumulate will also have a sizable content of dissipative dark matter. 
This motivates searching for the gravitational imprints of the dark sector within regions of high baryonic content, 
such as the Sun.

Finally, 
we point out that extensive numerical simulations of structure formation in dissipative dark-sector models are needed. 
Numerical simulations would confirm the existence of compact objects in our dark sector or other dissipative dark-sector models, 
may identify interesting features that our simplified analysis cannot capture, such as the distribution of exotic compact objects within the galaxy, 
and would provide further motivation for experimental efforts aimed at uncovering the behavior of dark matter on small scales. 

As the search for dark matter continues without any evidence of non-gravitational interactions of the dark and visible sectors,
theoretical and experimental efforts towards understanding the particle nature of dark matter purely based on
cosmological and astronomical observations will become increasingly important. 
Such efforts are complementary to the direct and indirect-detection programs,
and the interplay between both research directions may finally uncover the composition of the dark sector.

\subsection*{Acknowledgements}
We would like to thank Will Farr, Marilena LoVerde, Rosalba Perna, Cristobal Petrovich, Neelima Sehgal, and Anja von der Linden 
for useful discussions.
JHC and RE are supported by DoE Grant DE-SC0017938.  
The work of DE was supported in part by the National Science Foundation grant PHY-1620628. CK is partially funded by the Danish National Research Foundation, grant number DNRF90, and by the Danish Council for Independent Research, grant number DFF 4181-00055.    
RE and DE thank the Kavli Institute for Theoretical Physics for their hospitality.  
DE also thanks the Galileo Galilei Institute for Theoretical Physics for the hospitality, 
and the INFN for partial support during the completion of this work. 

\appendix

\section{Dark photon dark matter}
\label{app:one}

In this appendix we show the regions of dark-photon mass and temperature consistent with limits from warm dark matter and the number of effective relativistic degrees of freedom $N_{\textrm{eff}}$, 
if dark photons are the main constituent of the dark matter. 
The results are presented in Fig.~\ref{fig:eight}, as a function of the dark-photon mass and temperature at nucleosynthesis.
In the figure, 
the blue region is excluded by $\Delta N_{\textrm{eff}}$, Eq.~\eqref{eq:neff}.
The red region is excluded if dark photons are $100\%$ of the dark matter by the ``warm dark matter'' limit.
We crudely estimate the warm dark matter limit by requiring the dark photon free streaming length to be below $200\, \textrm{kpc}$ for consistency with structure on those scales~\cite{Gorbunov:2011zzc}. 
In the gray region the dark-photon thermal relic density Eq.~\eqref{eq:mgammadensity} is larger than the dark-matter density so the Universe overcloses. 
Along the gray line, a dark photon is $100\%$ of dark matter if the relic density is given by Eq.~\eqref{eq:mgammadensity}  (\textrm{i.e.}, if there is no further dark-photon production or depletion after they decouple relativistically).
In this case, 
from the figure we conclude that a thermal relic dark photon can be the dark matter if its mass is heavier than $m_{\gamma_D} \gtrsim 0.3 \, \textrm{keV}$.
\begin{figure} [t!]
\begin{center}
\includegraphics[width=8cm]{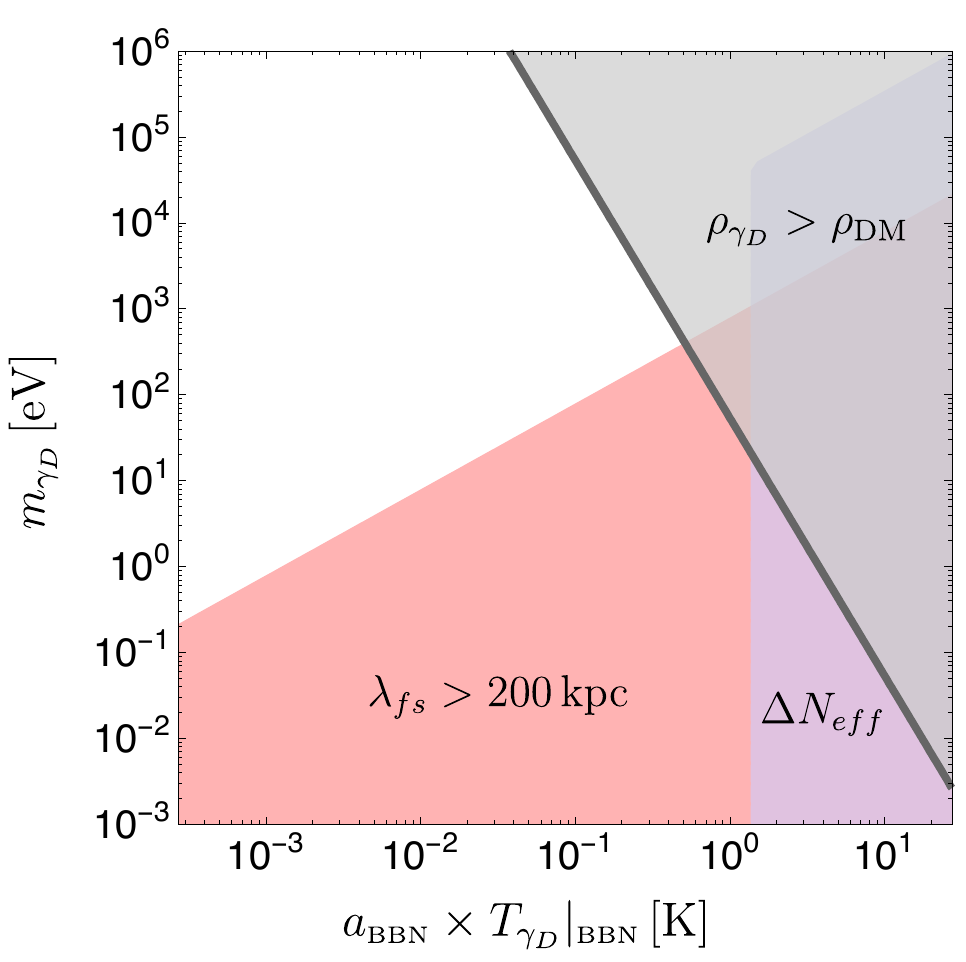}
\end{center}
\caption{Dark photon constraints from overclosure (gray), 
$\Delta N_{\textrm{eff}}$ (blue) as a function of the dark photon mass and temperature at nucleosynthesis.
In red we also present the constraints if the dark photons are $100\%$ of dark matter.}
\label{fig:eight}
\end{figure}
%%%%%%%%%%
% END FIGURE
%%%%%%%%%%

\section{Inverse Brems\-strahlung mean free path}
\label{app:two}
In this appendix we calculate the mean free-path for dark photon absorption. 
We assume thermal equilibrium throughout. 
The strategy is to obtain first the Brems\-strahlung rate,
then relate it to the inverse Bremsstrahlung rate using detailed balance,
and finally obtain the absorption mean free path directly from
\begin{equation}
\ell_{\gamma_D}^{\textrm{abs}} (\omega_{\gamma_D})
\equiv
\Gamma_{\textrm{IBS}}^{-1}
\quad .
\label{eq:libsdef}
\end{equation}

The dark-photon Brems\-strahlung rate for a dark-electron gas in thermal equilibrium is~\cite{Weldon:1983jn}
\begin{eqnarray}
\nonumber
\Gamma_{\textrm{BS}}(\omega_{\gamma_D})
&=&
\frac{1}{2 \omega_{\gamma_D}}
\int 
dp_1 dp_2
\,
dp_1' dp_2'
~n_1' n_2'
\\
&&
\,
\abs{\mathcal{M}(1',2' \rightarrow 1,2, \gamma_D)}^2
~
(2\pi)^4
\,
\delta^4
(
\,
p_{\gamma_{D}}+\sum_i (p_i-p_i')
\,
)
\quad ,
\end{eqnarray}
where $i=1,2$, $dp=d^3\mathbf{p}/2 E_p (2\pi)^3$, $n_{1,2}'$ are Fermi distribution functions, and we work in the non-degenerate limit, 
so we neglect Pauli blocking factors. 
The Brems\-strahlung rate may be rewritten in terms of the thermally weighted Brems\-strahlung differential cross section, 
which has been calculated already in the literature for a massless photon~\cite{Haug:1975bk2}. 
For relativistic dark-photon emission, we may neglect the dark-photon mass and use the result in~\cite{Haug:1975bk2}.
In this case, 
the Bremsstrahlung rate is 
\begin{equation}
\Gamma_{\textrm{BS}}
(\omega_{\gamma_D})
=
\frac{\pi^2}{\omega_{\gamma_D}^2}
\bigg<
\,
\frac{d\sigma_{\textrm{BS}}}{d\omega_{\gamma_D}}
v_{ee}
\,
\bigg>
\quad ,
\label{eq:GammaBS}
\end{equation}
where thermally-weighted cross section times electron relative velocity $v_{ee}$ is given by~\cite{Haug:1975bk2} 
\begin{eqnarray}
\nonumber
\bigg<
\,
\frac{d\sigma_{\textrm{BS}}}{d\omega_{\gamma_D}}
v_{ee}
\,
\bigg>
&\equiv &
\frac{\omega_{\gamma_D}}
{\pi^2}
~
\int 
dp_1 dp_2
\,
dp_1' dp_2'
~n_1' n_2'
\,
\abs{\mathcal{M}(1',2' \rightarrow 1,2, \gamma_D)}^2
~
(2\pi)^4
\,
\delta^4
(
\,
p_{\gamma_{D}}+\sum_i (p_i-p_i')
\,
)
\\
&=&
\frac{16\pi \alpha_D^3 n_{e_D}^2}
{15 m_{e_D}^3}
\,
\bigg(
\pi
\frac{T_{e_D}}
{m_{e_D}}
\bigg)^{-3/2}
\,
\frac{\omega_{\gamma_D}}
{m_{e_D}}
\,
\int_0^1
dx
\,
\frac{F(x)}
{x^3}
e^{-\omega_{\gamma_D}/T_{e_D}x}
\quad ,
\label{eq:dsigmaT}
\end{eqnarray}
where the function $F(x)$ is defined as
\begin{equation}
F(x)
\equiv
\bigg[
17
-
\frac{3x^2}
{(2-x)^2}
\bigg]
\sqrt{1-x}
+
\frac{12(2-x)^4
-
7x^2
(2-x)^2
-3x^4}
{(2-x)^3}
\,
\mathrm{ln}
\bigg[
\frac{1+\sqrt{1-x}}
{\sqrt{x}}
\bigg]
\quad .
\label{eq:Fx}
\end{equation}

The inverse Brems\-strahlung rate may now be related to the Brems\-strahlung rate via detailed balance~\cite{Weldon:1983jn}
\begin{equation}
\Gamma_{\textrm{IBS}}(\omega_{\gamma_D})
=
e^{\omega_{\gamma_D}/T_{e_D}}
\Gamma_{\textrm{BS}}(\omega_{\gamma_D})
\quad .
\label{eq:detailedbalance}
\end{equation}
Using Eq.~\eqref{eq:GammaBS} and~\eqref{eq:detailedbalance} in Eq.~\eqref{eq:libsdef}, 
we obtain the dark photon absorption mean free path
\begin{eqnarray}
\nonumber
\ell_{\gamma_D}^{\textrm{abs}} (\omega_{\gamma_D})
&\equiv&
\Gamma_{\textrm{IBS}}^{-1}
\\
&=&
\bigg[
e^{\omega_{\gamma_D}/T_{e_D}}
\frac{\pi^2}{\omega_{\gamma_D}^2}
\bigg<
\,
\frac{d\sigma_{\textrm{BS}}}{d\omega_{\gamma_D}}
v_{ee}
\,
\bigg>
\bigg]^{-1}
\nonumber 
\\
&=&
\bigg[
\frac{16\pi^3 \alpha_D^3 n_{e_D}^2}
{15 m_{e_D}^4 \omega_{\gamma_D}}
\,
\bigg(
\pi
\frac{T_{e_D}}
{m_{e_D}}
\bigg)^{-3/2}
\int_0^1
dx
\,
\frac{F(x)}
{x^3}
e^{\omega_{\gamma_D}/T_{e_D}(1-x)}
\bigg]^{-1}
\quad ,
\nonumber
\\
\label{eq:libs2}
\end{eqnarray}
where in the last equality we made use of Eq.~\eqref{eq:dsigmaT}.
In our dark-electron halo, 
dark photons are emitted with a typical energy $\sim T_{e_D}$. 
Using $\omega_{\gamma_D}= T_{e_D}$ in Eq.~\eqref{eq:libs2},we find 
\begin{equation}
\ell_{\gamma_D}^{\textrm{abs}}
=
8.3 \times 10^{-3}
~
\frac{({m_{e_D} T_{e_D}})^{5/2}}
{n_{e_D}^2 \alpha_D^3}
\quad .
\end{equation}
Equivalently, 
in terms of the dark-electron mass density, $\rho_{e_D} = n_{e_D} m_{e_D} $, the absorption mean-free path is
\begin{equation}
\ell_{\gamma_D}^{\textrm{abs}}
=
8.3 \times 10^{-3}
~
\frac{({m_{e_D}^{9} T_{e_D}^5})^{1/2}}
{\rho_{e_D}^2 \alpha_D^3}
\quad .
\end{equation}

\begin{figure} [t!]
\begin{center}
\includegraphics[width=8cm]{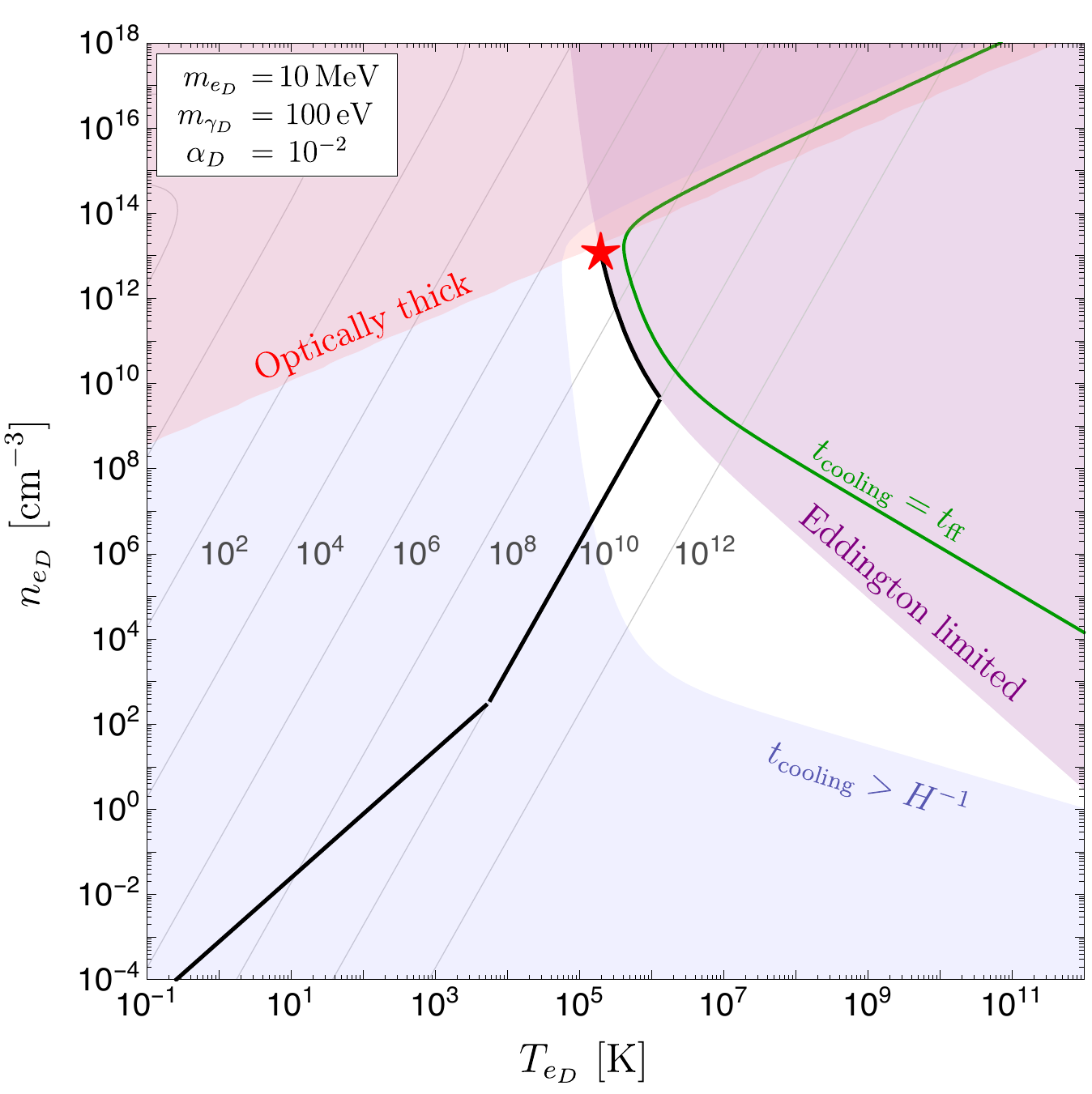}
\end{center}
\caption{Same as Fig.~\ref{fig:three} but including the radiation (Eddington) force. We shade in purple the regions of super-Eddington luminosity.
We also choose $m_{\gamma_D}=100 \, \textrm{eV}$, $m_{\gamma_D}=10 \, \textrm{MeV}$ and $\alpha_D=1/100$.}
\label{fig:nine}
\end{figure}

\section{Inclusion of Radiation Pressure for Halo Fragmentation}
\label{app:C}

In this appendix we include the effects of Eddington radiation pressure in the calculation of the minimal fragments. 
We include these effects by assuming that after reaching the Eddington limit, 
the dark-electron clumps virialize and contract in a nearly-virialized state by following the Eddington limit line Eq.~\eqref{eq:eddingtonlimit}.
This is illustrated in Fig.~\ref{fig:nine}.
Note that the Eddington limit line closely follows the ``cooling-equals heating'' line depicted in green. 
This is due to the fact that the collapse trajectory reaches both the Eddington limit line and the ``cooling-equals heating'' line when the cooling timescale becomes comparable to the gravitational dynamical timescale.
As a consequence, and as can be seen from Fig.~\ref{fig:nine},
the size of the minimal fragments obtained by following the Eddington or ``cooling-equals heating'' lines are similar. 
This is more comprehensively illustrated in Fig.~\ref{fig:ten}, 
where we present in dashed lines the mass of the minimal fragments including Eddington radiation pressure,
and in solid black without the inclusion of radiation pressure (i.e., as in Fig.~\ref{fig:five}).
We see that both results are quantitatively similar.

%%%%%%%%%%
% BEGIN FIGURE
%%%%%%%%%%
\begin{figure} [!]
\begin{center}
\begin{minipage}[c]{0.55\textwidth}
\includegraphics[width=8cm]{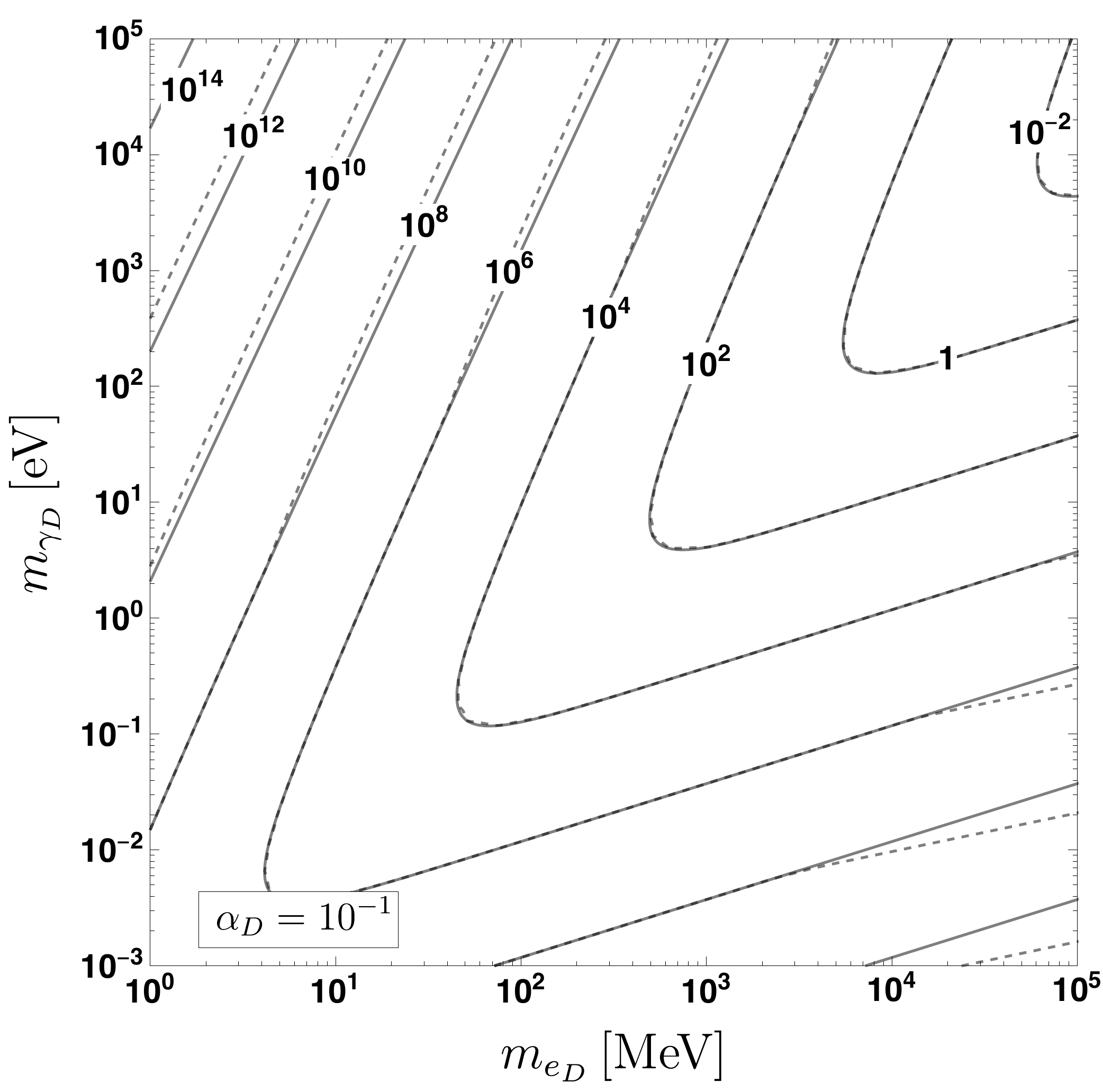}
\includegraphics[width=8cm]{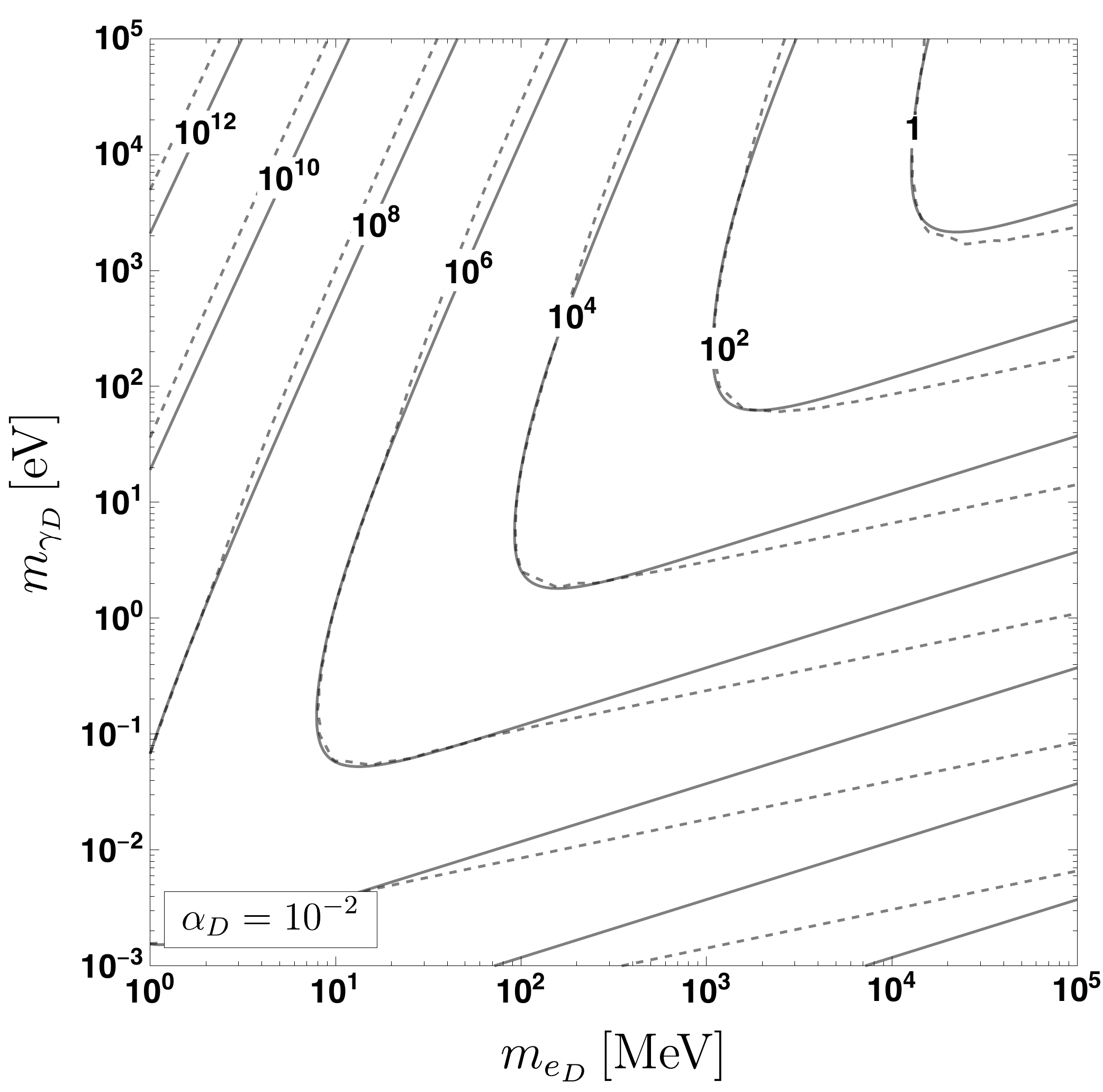}
\includegraphics[width=8cm]{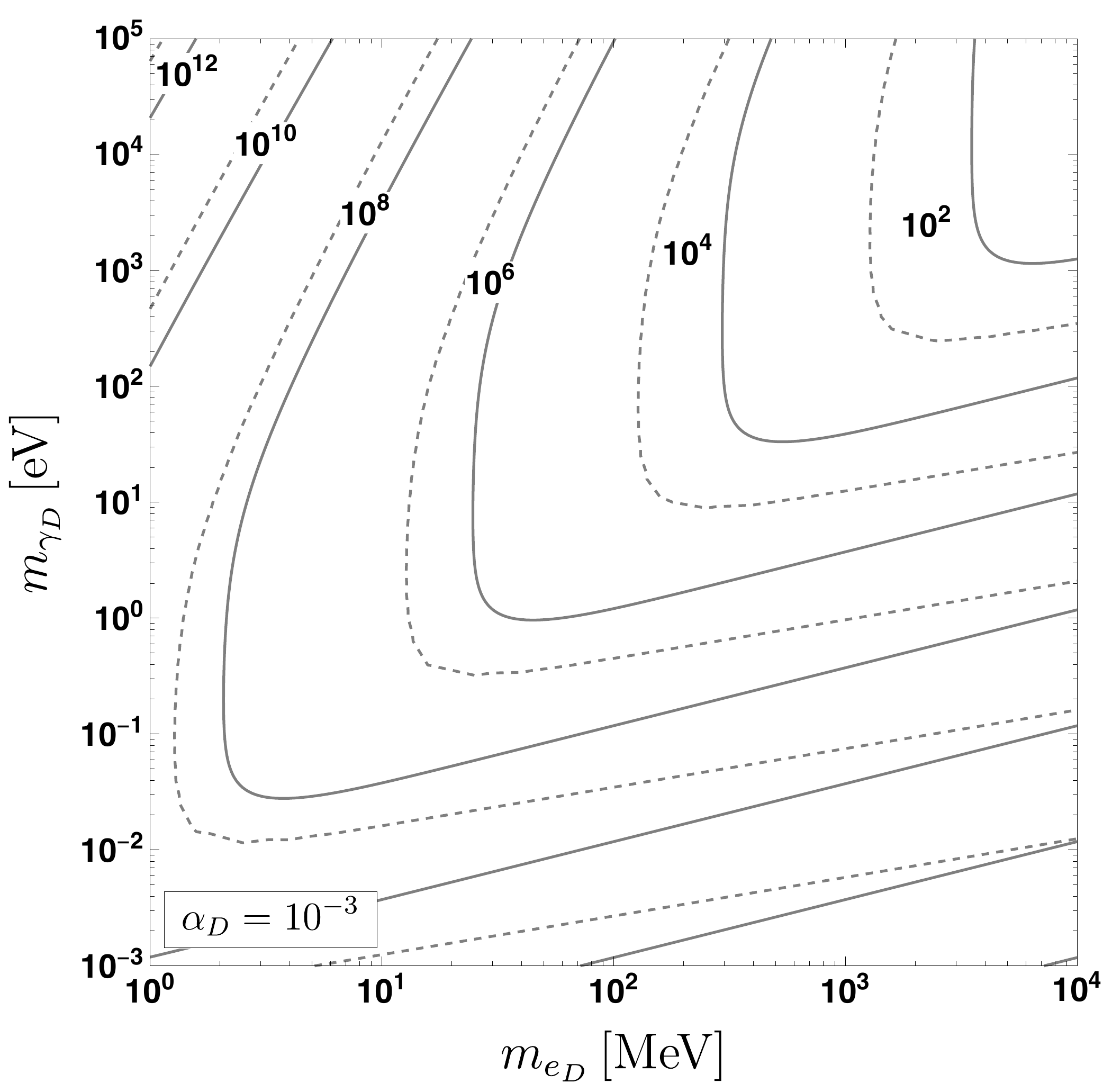}
  \end{minipage}\hfill
  \begin{minipage}[c]{0.45\textwidth}
    \caption{
Same as Fig.~\ref{fig:five} but showing in dashed lines the results obtained by including the Eddington force and for $\alpha_D=10^{-1}$ (top), $\alpha_D=10^{-2}$ (middle) and $\alpha_D=10^{-3}$ (bottom). By comparison we retain in solid black lines the results without the inclusion of the Eddington force.
    } 
    \label{fig:ten}
     \end{minipage}
\end{center}
\end{figure}
%%%%%%%%%%
% END FIGURE
%%%%%%%%%%

 %\FloatBarrier 
 
\section{Fragment compactness}
\label{app:three}
In this appendix we estimate the compactness of a homogeneous dark-electron gas sphere stabilized by the dark-photon repulsive force.
The dark-electron cloud is stabilized if its mass equals the Jeans mass,
 %%%
\begin{equation}
M
=
m_J\,,
\label{eq:stabilitycondition}
\end{equation}
%%%
where the Jeans mass due to the dark-photon pressure is 
\begin{eqnarray}
\nonumber
m_J &\equiv&
\frac{\pi}{6}
c_s^3
\bigg(
\frac{\pi}
{\rho G}
\bigg)
^{3/2}
n_{e_D} m_{e_D}
\\
&=&
\frac{4 \pi^4 n_{e_D}}
{3 m_{e_D}^2 m_{\gamma_D}^3}
\bigg(
\frac{30 \alpha_D^{3}}{G^3}
\bigg)^{1/2}\,,
\label{eq:Jeansmasscoulomb}
\end{eqnarray}
where in the second equality we used Eq.~\eqref{eq:idealgascs} assuming that the dark-photon force dominates over the kinetic pressure.
The dark-electron density for a homogeneous sphere is simply
\begin{equation}
n_{e_D}
=
\frac{3 M}
{4 \pi m_{e_D} R^3}
\quad ,
\label{eq:density}
\end{equation}
%%%
where $M$ and $R$ are the mass and radius of the dark electron gas sphere.
Using Eq.~\eqref{eq:density} in Eq.~\eqref{eq:Jeansmasscoulomb} and solving the stability condition Eq.~\eqref{eq:stabilitycondition}, we find the dark-electron sphere radius
\begin{equation}
R
=
\frac{\pi}
{m_{e_D} m_{\gamma_D}}
\bigg(
\frac{\alpha_D}
{G}
\bigg)^{1/2}
\quad ,
\end{equation}
so the compactness, 
defined in Eq.~\eqref{eq:compactnessdef} is
\begin{equation}
C
=
\frac{ m_{e_D} m_{\gamma_D} M}
{\pi}
\bigg(
\frac{G^3}
{ \alpha_D}
\bigg)^{1/2}
\quad .
\end{equation}

\FloatBarrier

\bibliography{treeA_bib}

\end{document}